\documentclass[notitlepage,prd,preprintnumbers,longbibliography,showpacs]{revtex4-1}
\usepackage[%
colorlinks=true,
urlcolor=blue,
linkcolor=blue,
citecolor=blue
]{hyperref}
\usepackage{slashed,color,amsmath,amssymb,mathrsfs}
\usepackage{cancel}
\usepackage{autobreak}
\usepackage{graphicx}
\usepackage{hyperref}
\usepackage{longtable}
\usepackage{array}
\usepackage{accents}
\usepackage{tensor}
\usepackage{longtable}
\usepackage{booktabs}
\usepackage{multirow}
\usepackage{bm}
\graphicspath{{figures/}}

\allowdisplaybreaks

\newcommand{\la}{\langle}
\newcommand{\ra}{\rangle}

\newcommand{\chip}{\chi_+}

\newcommand{\chim}{\chi_-}

\newcommand{\gamf}{\gamma_5}

\newcommand{\Bb}{\overline{B}}

\newcommand{\Tb}{\overline{T}}

\newcommand{\psib}{\overline{\psi}}

\begin{document}

\title[]{Chiral Lagrangians for spin-$\frac{1}{2}$ and spin-$\frac{3}{2}$ doubly charmed baryons }

\author{Hao Liu$^{1}$}
\author{Yuan-He Zou$^{1}$}
\author{Yan-Rui Liu$^2$}\email{yrliu@sdu.edu.cn}
\author{Shao-Zhou Jiang$^{1}$}\email{jsz@gxu.edu.cn}

\affiliation{$^{1}$ Key Laboratory for Relativistic Astrophysics, School of Physical Science and Technology, Guangxi University, Nanning 530004, People's Republic of China\\
$^2$School of Physics, Shandong University, Jinan 250100, People's Republic of China}
\date{\today}
\begin{abstract}

The relativistic chiral Lagrangians for both spin-$\frac{1}{2}$ and spin-$\frac{3}{2}$ doubly charmed baryons are constructed up to the order $\mathcal{O}(p^{4})$. From $\mathcal{O}(p^{2})$ to $\mathcal{O}(p^{4})$, there are 19, 74, and 452 independent terms in the two-flavor case and 25, 112, and 864 independent terms in the three-flavor case. The chiral Lagrangians in the heavy diquark limit are also obtained. From $\mathcal{O}(p^{2})$ to $\mathcal{O}(p^{4})$, there are 7, 23, and 118 independent terms in the two-flavor case and 8, 31, and 189 independent terms in the three-flavor case. We present the low-energy constant relations between the relativistic case and the case in the heavy diquark limit up to the order $\mathcal{O}(p^{3})$. With the heavy diquark-antiquark symmetry, the low-energy constant relations between the doubly charmed baryon case and the heavy-light meson case are also obtained up to the order $\mathcal{O}(p^{3})$.
\end{abstract}

\maketitle

%%%%%%%%%%%%%%%%%%%%%%%%%%%%%%%%%%%%%%%%%%%%%%
\section{Introduction}\label{sec:level1a}
%%%%%%%%%%%%%%%%%%%%%%%%%%%%%%%%%%%%%%%%%%%%%%

Heavy quark baryons play an important role in hadron physics and the study of their properties may deepen our understanding of QCD. Their spectra are tightly related to the non-perturbative effects of QCD \cite{Chen:2016spr}. Until now, all types of singly heavy baryons have been observed \cite{ParticleDataGroup:2022pth} while only one doubly heavy baryon $\Xi_{cc}^{++}$ is reported by two experiments, SELEX \cite{Russ:2002bw} and LHCb \cite{LHCb:2017iph}. Although the SELEX Collaboration claimed the observation of $\Xi_{cc}^+$ with mass $3519\pm1$ MeV \cite{SELEX:2002wqn}, other measurements from FOCUS \cite{Ratti:2003ez}, BaBar \cite{BaBar:2006bab}, Belle \cite{Belle:2006edu}, and LHCb \cite{LHCb:2013hvt} did not confirm this result. For the mass of the $\Xi_{cc}$ baryon, the LHCb Collaboration obtains a value about 100 MeV higher than the SELEX Collaboration. The puzzle for the mass inconsistency needs more studies to solve.

The observation of $\Xi_{cc}^{++}$ in the LHCb experiment inspired lots of discussions on the properties about doubly charmed baryons. Various theoretical approaches were adopted in the discussions, such as chiral perturbation theory (ChPT), QCD sum rule, lattice QCD, and heavy quark effective theory. Some properties of spin-$\frac{1}{2}$ and spin-$\frac{3}{2}$ doubly charmed baryons have been explored extensively, like the mass spectra \cite{Lu:2017meb,Weng:2018mmf,Yan:2018zdt,Kiselev:2017eic,Kerbikov:2017pau,Karliner:2018hos,Yan:2018zdt,Gutierrez-Guerrero:2019uwa,Mathur:2018rwu,Yao:2018ifh,Wang:2017qvg,Yu:2018com,Li:2019ekr,Sun:2014aya}, electromagnetic properties \cite{Meng:2017dni,Li:2017pxa,Liu:2018euh,HillerBlin:2018gjw,Bahtiyar:2018vub,Ozdem:2018uue,Li:2020uok,Li:2017cfz,Shi:2021kmm,Simonis:2018rld}, weak and strong decay properties \cite{Wang:2017azm,Zhao:2018mrg,Xiao:2017udy,Xiao:2017udy,Gutsche:2017hux,Gutsche:2019iac,Gutsche:2019wgu,Han:2021azw,Geng:2017mxn,Shi:2017dto,Zhang:2018llc,Shi:2019fph,Hu:2019bqj,Wang:2017mqp,Zhao:2018zcb,Zhao:2018mrg,Xing:2018lre,Hu:2020mxk,Sharma:2017txj,Dhir:2018twm,Zou:2019kzq,Hu:2017dzi}, and so on. Here, we would like to discuss the chiral Lagrangians in studying the properties of doubly charmed baryons in ChPT.

As an effective theory of QCD, ChPT provides an alternative approach to deal with strong interactions over long distances \cite{Weinberg:1978kz,Gasser:1983yg,Gasser:1984gg}. It is a powerful tool in studying low-energy processes involving pions. In ChPT, the eight low-lying pseudoscalar mesons are treated as the Goldstone bosons generated by spontaneous breaking of chiral symmetry of QCD. These mesons instead of the complicated quarks and gluons mediate the interactions between hadrons. Only the pseudoscalar mesons were initially incorporated in the framework. Later, various ChPTs including baryons were developed \cite{Krause:1990xc,Jenkins:1991es,Hemmert:1997ye,Yan:1992gz,Wise:1992hn}.

In ChPT, the chirally invariant Lagrangian is the most basic premise for practical applications. For a theoretical study, in principle, more higher order chiral Lagrangians are involved, and therefore more precise results would be obtained. However, with the increase of chiral order, the number of independent interaction terms and the corresponding unknown low-energy constants (LECs) would become considerable. These LECs cannot be determined by ChPT itself. Even worse, they are also hard to be obtained in other ways. This makes it quite difficult to work with high-order chiral Lagrangians. In order to solve this problem to some extent, available approximate symmetries, e.g. heavy quark symmetry and heavy diquark-antiquark symmetry \cite{Burdman:1992gh,Meng:2018zbl}, could be introduced to constraint the number of LECs or to give some approximate LEC relations. The present study will involve such symmetries.

Of the six flavor quarks in QCD, three light quarks are below the scale of the chiral symmetry breaking, $\Lambda_\chi\sim 1$ GeV. The charm quark is above $\Lambda_\chi$ and its mass $M_c$ is much larger than those of light quarks ($M_q$) as well as the scale $\Lambda_{\mathrm{QCD}}\sim 200$ MeV. Then, the ratios $M_q/M_c$ and $\Lambda_{\mathrm{QCD}}/M_c$ are both considered small enough. In the extreme limit $M_c\to\infty$, both $M_q/M_c$ and $\Lambda_{\mathrm{QCD}}/M_c$ could be ignored and the heavy quark symmetry of QCD appears \cite{Isgur:1989vq,Georgi:1990ak}. In this limit, the dynamics of the considered hadron may not be affected by the change of the heavy quark flavor (spin), which corresponds to the heavy quark flavor (spin) symmetry. For the doubly charmed baryons we are considering, one may assume that the two charm quarks form a compact heavy diquark with spin=1. We will use {\it heavy diquark limit} to denote the case that the mass of the diquark does not affect the dynamics of the doubly charmed baryon \cite{Meng:2018zbl}.

With the chiral Lagrangians, many properties of doubly charmed baryons have been studied in the literature, e.g. their masses \cite{Sun:2014aya,Sun:2016wzh,Yao:2018ifh}, magnetic moments \cite{Meng:2017dni,Liu:2018euh,Shi:2021kmm}, and electromagnetic form factors \cite{Li:2017pxa,HillerBlin:2018gjw}. At present, most studies involve the low order Lagrangians. Parts of high order chiral Lagrangians are given in a few studies on some specific problems \cite{Sun:2014aya,Meng:2017dni,Li:2017pxa,Liu:2018euh,HillerBlin:2018gjw,Li:2020uok}. Recently, the chiral Lagrangian for spin-$\frac{1}{2}$ doubly charmed baryons in the three-flavor case has been constructed up to the order $\mathcal{O}(p^{4})$ \cite{Qiu:2020omj}. In the present work, for considerations to complete the necessary ingredient of one-loop level investigations, to check convergence of chiral expansion better, to motivate future LEC studies, and so on, we construct the relativistic chiral Lagrangian involving both spin-$\frac{1}{2}$ and spin-$\frac{3}{2}$ doubly charmed baryons up to the order $\mathcal{O}(p^{4})$. In principle, when one determines
the values of LECs from experimental data, all the operators should be independent. Otherwise, the overfitting problem would appear. The Lagrangians given in the present study will be minimal and the terms will be independent. Besides the relativistic case, we will also consider Lagrangians in the heavy diquark case. In that case, the spin-1/2 and spin-3/2 baryons can be put into a superfield and some LEC relations can be obtained.

This paper is organized as follows. Section \ref{sec:level1b} introduces the building blocks in constructing the chiral Lagrangians for doubly charmed baryons. In Sec. \ref{sec:level1c}, the properties of the building blocks, the needed linear relations, and the procedure to get the relativistic and non-relativistic forms of Lagrangians are given. In Sec. \ref{sec:level1e}, we discuss how to find constraints on LECs with the heavy quark symmetry from different Lagrangians. Section \ref{sec:level1f} shows our results and some discussions. Sec. \ref{sec:level1g} is a short summary.

%%%%%%%%%%%%%%%%%%%%%%%%%%%%%%%%%%%%%%%%%%%%%%%%%%%%%%%%%%%%%%%%%%%%%%%
\section{Definitions and building blocks}\label{sec:level1b}
%%%%%%%%%%%%%%%%%%%%%%%%%%%%%%%%%%%%%%%%%%%%%%%%%%%%%%%%%%%%%%%%%%%%%%%

This section presents the definitions related with pseudoscalar Goldstone boson fields, external sources, and doubly charmed baryons in both relativistic and non-relativistic forms. One may find more relevant details in Refs. \cite{Sun:2014aya,Meng:2017dni,Gasser:1983yg,Gasser:1984gg,Gasser:1987rb,Fearing:1994ga,Bijnens:1999sh,Bijnens:2001bb,Cata:2007ns,Fettes:2000gb,Oller:2006yh,Jiang:2016vax}.

\subsection{Goldstone boson fields and external sources}\label{bbb}
Considering the lightest $N_{f}$-flavor quarks ($N_{f}=2$ or $3$), one may write the related QCD Lagrangian with external sources as
\begin{eqnarray}
\mathcal{L}=\mathcal{L}_{\mathrm{QCD}}^{0}+\bar{q}\left(\slashed{v}+\slashed{a}\gamma_{5}-s+i p \gamma_{5}\right) q,
\end{eqnarray}
where $\mathcal{L}_{\mathrm{QCD}}^{0}$ is the original QCD Lagrangian and $q$ denotes the light quark fields. Here, $s$, $p$, $v^{\mu}$, and $a^{\mu}$ are scalar, pseudoscalar, vector, and axial-vector external sources, respectively. In this work, the tensor source and the $\theta$ term are not considered. The external source $a^{\mu}$ is always traceless, but $v^{\mu}$ is not. To consider the electromagnetic interactions, $v^\mu$ is taken to be traceable in the relativistic case. In the heavy diquark limit, it is traceable (traceless) in the $SU(2)$ ($SU(3)$) case because the heavy diquark is just a spectator and the electromagnetic interaction is mainly determined by the light quark.

When the light quarks are taken to be massless, QCD exhibits a global $SU(N_{f})_{L} \times SU(N_{f})_{R}$ chiral symmetry, but it spontaneously breaks into $SU(N_{f})_{V}$ because of the nonvanishing quark condensate $\langle\bar{q}q\rangle$. The generated $(N_f^2-1)$ Goldstone bosons are considered to be the lowest pseudoscalar pion mesons. These mesonic fields are collected in the nonlinear representation $u=\exp(i\pi^j\tau^j/2F_0)$ or $u=\exp(i\pi^j\lambda^j/2F_0)$ where $F_0$ is the pion decay constant in the chiral limit, $\tau^j$ ($j$=1,2,3) are the Pauli matrices, and $\lambda^j$ ($j$=1,2,$\cdots$,8) are the Gell-Mann matrices. As the basic ingredient of ChPT, $u$ has the chiral transformation behavior
\begin{eqnarray}
u \rightarrow g_{L} u h^{\dagger}=h u g_{R}^{\dagger},
\end{eqnarray}
where $g_{L}$ and $g_{R}$ are the group elements of $SU(N_{f})_{L}$ and $SU(N_{f})_{R}$, respectively, and $h$ is a compensator field related to the pion fields.

The necessary building blocks in constructing the chiral Lagrangians are combinations of the mesonic fields and the external sources. The conventional choice of their explicit forms reads
\begin{align}\label{buildingblocks}
u^{\mu} &=i\left\{u^{\dagger}\left(\partial^{\mu}-i r^{\mu}\right) u-u\left(\partial^{\mu}-i l^{\mu}\right) u^{\dagger}\right\} ,\notag\\
h^{\mu \nu} &=\nabla^{\mu} u^{\nu}+\nabla^{\nu} u^{\mu} ,\notag\\
f_{+}^{\mu \nu} &=u F_{L}^{\mu \nu} u^{\dagger}+u^{\dagger} F_{R}^{\mu \nu} u ,\notag\\
f_{-}^{\mu \nu} &=u F_{L}^{\mu \nu} u^{\dagger}-u^{\dagger} F_{R}^{\mu \nu} u=-\nabla^{\mu} u^{\nu}+\nabla^{\nu} u^{\mu},\notag\\
\chi_{\pm} &=u^{\dagger} \chi u^{\dagger} \pm u \chi^{\dagger} u ,
\end{align}
where $r^{\mu}=v^{\mu}+a^{\mu}$, $l^{\mu}=v^{\mu}-a^{\mu}$, $F_{R}^{\mu \nu}=\partial^{\mu} r^{\nu}-\partial^{\nu} r^{\mu}-i\left[r^{\mu}, r^{\nu}\right]$, $F_{L}^{\mu \nu}=\partial^{\mu} l^{\nu}-\partial^{\nu} l^{\mu}-i\left[l^{\mu},l^{\nu}\right]$, and  $\chi=2 B_{0}(s+i p)$ with $B_{0}$ being a constant related to the quark condensate. The covariant derivative of a building block $X$ is defined by
\begin{align}
&\nabla^{\mu}X=\partial^{\mu} X+\left[\Gamma^{\mu}, X\right], \label{deri}\\
&\Gamma^{\mu}=\frac{1}{2}\left\{u^{\dagger}\left(\partial^{\mu}-i r^{\mu}\right) u+u\left(\partial^{\mu}-i l^{\mu}\right) u^{\dagger}\right\}.
\end{align}
The chiral dimension of $\nabla^\mu$ is one in ChPT. All the defined building blocks have the same behavior under the chiral rotation $R$,
\begin{eqnarray}
X \stackrel{R}{\longrightarrow} X^{\prime}=h X h^{\dagger}.
\end{eqnarray}

\subsection{Doubly charmed baryons}

The spin of an $S$-wave doubly charmed baryon containing two charmed quarks and one light quark may be 1/2 or 3/2. Then the doubly charmed baryons belong to two triplets in the three-flavor case,
\begin{eqnarray}
B=\begin{pmatrix}
\Xi_{c c}^{++} \\
\Xi_{c c}^{+} \\
\Omega_{c c}^{+}
\end{pmatrix}, \quad T^{ \mu}=\begin{pmatrix}
\Xi_{c c}^{*++} \\
\Xi_{c c}^{*+} \\
\Omega_{c c}^{*+}
\end{pmatrix},
\end{eqnarray}
where $B$ and $T^{\mu}$ denote the spin-$\frac{1}{2}$ and spin-$\frac{3}{2}$ fields, respectively. Only the $\Xi_{cc}^{(*)}$ states are needed if one considers the two-flavor case. The covariant derivative of a baryon field and its Dirac conjugate are
\begin{align}
D^{\mu} \tilde{B}=\left(\partial^{\mu}+\Gamma^{\mu}\right) \tilde{B},\qquad
D^{\mu} \bar{\tilde{B}}=\bar{\tilde{B}}\left(\overleftarrow{\partial}^{\mu}+\Gamma^{\mu^{\dagger}}\right),
\end{align}
where $\tilde{B}=B$ or $T^{ \nu}$. If more than one covariant derivatives act on $\tilde{B}$, a totally symmetrical derivative is introduced
\begin{align}
D^{\mu \nu \cdots \rho} &\equiv \frac{1}{n !}(\underbrace{D^{\mu} D^{\nu} \cdots D^{\rho}}_{n}+\text{full permutation of $D$’s}).
\end{align}
Any antisymmetrical derivative is related to the higher order contributions \cite{Fettes:2000gb}. The chiral rotations of the baryon fields (and their derivatives) are
\begin{eqnarray}
\tilde{B} \stackrel{R}{\longrightarrow} \tilde{B}^{\prime}=h\tilde{B}, \qquad \bar{\tilde{B}} \stackrel{R}{\longrightarrow} {\bar{\tilde{B}}^{\prime}}=\bar{\tilde{B}} h^{\dagger}.
\end{eqnarray}

In the heavy diquark limit, a doubly charmed baryon with velocity $v^{\mu}$ \footnote{From now on, $v^\mu$ will no longer denote vector external source.} can be simply separated into a heavy diquark component and a light degree of freedom component. The diquark behaves just like a static color source for the light component and the suppression for the spin interaction between the diquark and the light component results in the degenerate $B$ and $T^\mu$. In this situation, the heavy diquark symmetry exists and one may put the spin-$\frac{1}{2}$ and spin-$\frac{3}{2}$ baryon fields into a superfield $\psi^{\mu}$ \cite{Meng:2018zbl},
\begin{eqnarray}
\psi^{\mu} &=T^{\mu}+\sqrt{\frac{1}{3}}\left(\gamma^{\mu}+v^{\mu}\right) \gamma^{5} B ,\label{psi}\\
\bar{\psi}^{\mu} &=\bar{T}^{\mu}-\sqrt{\frac{1}{3}} \bar{B} \gamma^{5}\left(\gamma^{\mu}+v^{\mu}\right).
\end{eqnarray}
Here, $B$ and $T^{\mu}$ as non-relativistic fields only contain the annihilation operators. It is obvious that the chiral rotation behaviors of $\psi^{\mu}$ and $\bar{\psi}^{\mu}$ are the same as $\tilde{B}$ and $\bar{\tilde{B}}$, respectively. In order to modify the energy measure, the superfield is scaled by $e^{-iMv\cdot x}$ with $M$ being the doubly charmed baryon mass. Hence, in the heavy diquark limit, the covariant derivative on matter field becomes
\begin{align}
D^{\mu} \psi^{\nu}(x)=-i M v^{\mu} \psi^{\nu}(x).
\end{align}

%%%%%%%%%%%%%%%%%%%%%%%%%%%%%%%%%%%%%%%%%%%%%%%%%%%%%%%%%%%%%%%%%%%%%%%%%%%%%%%%%%%%%%%%%%%%%
\section{Construction of Chiral Lagrangians for doubly Charmed Baryons}\label{sec:level1c}
%%%%%%%%%%%%%%%%%%%%%%%%%%%%%%%%%%%%%%%%%%%%%%%%%%%%%%%%%%%%%%%%%%%%%%%%%%%%%%%%%%%%%%%%%%%%%

The basic procedure to construct the chirally invariant Lagrangians for doubly charmed baryons is as follows. First, the structures of the chiral Lagrangians are introduced. Second, the necessary properties of the building blocks are given. One constructs all possible terms based on such properties. Third, all linear relations to reduce the number of Lagrangian terms are listed. Finally, the linearly dependent terms would be eliminated and the independent ones are retained.

\subsection{Structures of Lagrangians}

The relativistic chiral Lagrangian ${\cal L}$ for doubly charmed baryons contains terms involving only spin-$\frac{1}{2}$ baryons, those involving only spin-$\frac{3}{2}$ baryons, and those involving both spin-$\frac{1}{2}$ and spin-$\frac{3}{2}$ baryons. We use $\mathcal{L}_{B B}$, $\mathcal{L}_{TT}$, and $\mathcal{L}_{B T}$ to represent these three sectors of interaction terms, respectively. That is,
\begin{align}
\mathcal{L}&=\mathcal{L}_{B B}+\mathcal{L}_{TT}+\mathcal{L}_{B T}\notag\\
&= \sum_{n} E_{n} \bar{B} \cdots B+\sum_{m} E_{m} \bar{T} \cdots T
+\sum_{p} E_{p}\left(\bar{B} \cdots T+\bar{T} \cdots B\right),\label{eq:rlag}
\end{align}
where $E_{n}$, $E_{m}$, and $E_{p}$ are LECs and the symbol “$\cdots$” represents the allowed combinations of building blocks. If the trace of a combination $X$ in the flavor space is necessary, it will be denoted as $\langle X\rangle$. To keep the hermiticity of $\mathcal{L}$, there would exist some appropriate coefficients $\pm i$ in “$\cdots$”.

In the heavy diquark limit, the Lagrangian has the structure
\begin{eqnarray}\label{eq:hlag}
\mathcal{L}_{\mathrm{H}}=\sum_{n} F_{n} \bar{\psi}^{\mu}\cdots\psi_{\mu}
\end{eqnarray}
where $F_{n}$ are LECs.

\subsection{Properties of building blocks}

\begin{table}[h]
\caption{\label{tab:table1}%
Chiral dimension (Dim), parity transformation ($P$), charge conjugation ($C$), and Hermiticity (H.c.) of the building blocks.}
\begin{ruledtabular}
\begin{tabular}{ccccc}
                     & \textrm{Dim} &       \textrm{P}       &          \textrm{C}          &   \textrm{H.c.}    \\
  \colrule
$u^{\mu}$ &      1       &       $-u^{\mu}$       &       $(u^{\mu})^{T}$        &     $u^{\mu}$      \\
     $h^{\mu\nu}$    &      2       &     $-h^{\mu\nu}$      &      $(h^{\mu\nu})^{T}$      &    $h^{\mu\nu}$    \\
     $\chi_{\pm}$    &      2       &    $\pm\chi_{\pm}$     &      $(\chi_{\pm})^{T}$      &  $\pm\chi_{\pm}$   \\
  $f_{\pm}^{\mu\nu}$ &      2       & $\pm f_{\pm}^{\mu\nu}$ & $\mp (f_{\pm}^{\mu\nu})^{T}$ & $f_{\pm}^{\mu\nu}$
\end{tabular}
\end{ruledtabular}
\end{table}

\begin{table}[h]
	\caption{\label{tab:table2}%
		Chiral dimension (Dim), parity transformation ($P$), charge conjugation($C$), and Hermiticity (H.c.) of the Clifford algebra, the velocity of doubly charmed baryons, Levi-Civita tensor, and the covariant derivatives acting on baryons. The subscripts $BB$, $TT$, and $BT$ represent $\mathcal{L}_{BB}$, $\mathcal{L}_{TT}$, and $\mathcal{L}_{B T}$, respectively.}
	\begin{ruledtabular}
		\begin{tabular}{cccccccc}
			&
			\textrm{Dim}&
			$\textrm{P}_{BB(TT)}$&
			$\textrm{C}_{BB(TT)}$&
			$\textrm{H.c.}_{BB(TT)}$&
			$\textrm{P}_{BT}$&
			$\textrm{C}_{BT}$&
			$\textrm{H.c.}_{BT}$\\
			\colrule
			1 & 0 & $+$ & $+$ & $+$ & $-$ & $+$ & $+$ \\
			$\gamma_{5}$ & 1 & $-$ & $+$ & $-$ & $+$ & $+$ & $-$ \\
			$\gamma^{\mu}$ & 0 & $+$ & $-$ & $+$ & $-$ & $-$ & $+$ \\
			$\gamma_{5}\gamma^{\mu}$ & 0 & $-$ & $+$ & $+$ & $+$ & $+$ & $+$\\
			$\sigma^{\mu\nu}$ & 0 & $+$ & $-$ & + & $-$ & $-$ & $+$\\
			$v^{\mu}$ & 0 & $+$ & $-$ & $+$ & $+$ & $-$ & $+$ \\
			$\epsilon^{\mu\nu\lambda\rho}$& 0 & $-$ & $+$ & $+$ & $+$ & $+$ & $+$\\
			$D^{\mu}\tilde{B}$ & 0& $+$ & $-$ & $-$ & $+$ & $+$ & $+$
		\end{tabular}
	\end{ruledtabular}
\end{table}

The chiral Lagrangian should be invariant under the parity transformation, charge conjugation, and hermitian conjugation. Table \ref{tab:table1} shows the transformation behaviors of the building blocks about mesons and external sources. Table \ref{tab:table2} presents those of the Clifford algebra, the velocity of doubly charmed baryons, Levi-Civita tensor, and the covariant derivatives acting on baryons. With these transformation properties, all the constructed terms would have the structures shown in Eqs. \eqref{eq:rlag} and \eqref{eq:hlag}. In Table \ref{tab:table2}, we only show the transformation signs. Some explanations are as follows. Under the parity transformation, one has
\begin{eqnarray}
B\stackrel{P}{\longrightarrow}\gamma_{0}B, \quad {T}_{\mu} \stackrel{P}{\longrightarrow} -\gamma_{0}T_{\mu}.
\end{eqnarray}
For convenience in the Lagrangian construction, we omit the minus sign in the second transformation and compensate it in the transformations of the five Clifford algebra elements and the antisymmetric tensor $\epsilon^{\mu\nu\lambda\rho}$. That is why we distinguish the transformations between $\mathcal{L}_{B T}$ and $\mathcal{L}_{BB(TT)}$ in Table \ref{tab:table2}. Moreover, we use a convention that the covariant derivatives in ${\cal L}_{BB}$ and ${\cal L}_{TT}$ of Eq. \eqref{eq:rlag} act on the right baryon field $\tilde{B}$, but those in ${\cal L}_{BT}$ act only on the spin-3/2 field ${T}_{\mu}$ or $\bar{T}_{\mu}$. Therefore, the properties of covariant derivatives in $\mathcal{L}_{BB(TT)}$ and $\mathcal{L}_{B T}$ may also be different. However, the convention difference has no impact on the construction of ${\cal L}_{BT}$. The related properties for $\mathcal{L}_{\mathrm{H}}$ in Eq. \eqref{eq:hlag} are the same as those for $\mathcal{L}_{TT}$.

With these tables, a complete set of interaction terms can be obtained. However, these constructed terms are not always linearly independent. The next subsection will list all possible linear relations with which the redundant terms can be eliminated.

\subsection{Linear relations}\label{sec:level2c2}
The linear relations in the construction of effective Lagrangians have been discussed widely. We here just give a short description of them. More details can be found in Refs. \cite{Fearing:1994ga,Qiu:2020omj,Bijnens:1999sh,Bijnens:2001bb,Fettes:2000gb,Jiang:2017yda}.

\subsubsection{Partial integration}

A derivative acting on any Lagrangian term does not affect the physics. Therefore, the covariant derivative acting on one baryon field can be changed to that acting on the other one, which induces a high-order difference, i.e.
\begin{align}
0 &\doteq \bar{B}  \overleftarrow{D}^{\nu} X B+\bar{B} X D^{\nu} B,\\
0 &\doteq \bar{T}^{\mu}  \overleftarrow{D}^{\nu} X {T}_{\mu}+\bar{T}^{\mu} X D^{\nu} {T}_{\mu},\\
0 &\doteq \bar{B} \overleftarrow{D}^{\nu} X {T}^{\mu}+\bar{B} X D^{\nu} {T}^{\mu},\\
0 &\doteq \bar{T}^{\mu} \overleftarrow{D}^{\nu} X B+\bar{T}^{\mu} X D^{\nu} B,
\end{align}
where $X$ denotes any possible combination of the building blocks and ``$\doteq$'' means that the higher order terms are ignored. We will use the first two relations to move the positions of covariant derivatives. The third and fourth relations will not give any new constraints in the present work.

\subsubsection{Schouten identity}

The Schouten identity involves the four-dimension Levi-Civita tensor,
\begin{eqnarray}
\epsilon^{\mu \nu \lambda \rho} A^{\sigma}-\epsilon^{\sigma \nu \lambda \rho} A^{\mu}-\epsilon^{\mu \sigma \lambda \rho} A^{\nu}-\epsilon^{\mu \nu \sigma \rho} A^{\lambda}
-\epsilon^{\mu \nu \lambda \sigma} A^{\rho}=0,
\end{eqnarray}
where $A$ can be any building block containing at least one Lorentz index. This relation exists because the five indices in the equation are antisymmetric, but the dimension of the spacetime is four.

\subsubsection{Equations of motions (EOMs)}

The equations of motions for the light pseudoscalar mesons and the doubly charmed baryons read
\begin{align}
&\nabla_{\mu} u^{\mu}\doteq \frac{i}{2}\left(\chi_{-}-\frac{1}{N_{f}}\left\langle\chi_{-}\right\rangle\right),\\
&\left(i \slashed{D}-m_{T}\right) T^{\mu} \doteq 0,\\
&\left(i \slashed{D}-m_{B}\right) B \doteq 0.
\end{align}
For the spin-3/2 fields, two subsidiary conditions are needed to eliminate the redundant degrees of freedom,
\begin{align}
&D_{\mu} T^{\mu} \doteq 0, \label{EOMh}\\
&\gamma_{\mu} T^{\mu} \doteq 0.\label{EOMinheavylimit}
\end{align}
In the heavy diquark limit, one has
\begin{align}
v_{\mu} \psi^{\mu}=0,\\
\slashed{v}\psi^{\mu}=\psi^{\mu}.
\end{align}

\subsubsection{Covariant derivatives and Bianchi identity}

The commutation of two covariant derivatives gives
\begin{align}
&\left[\nabla^{\mu}, \nabla^{\nu}\right] X=\left[\Gamma^{\mu \nu}, X\right],\\
&\Gamma^{\mu \nu}=\frac{1}{4}\left[u^{\mu}, u^{\nu}\right]-\frac{i}{2} f_{+}^{\mu \nu}.
\end{align}
The Bianchi identity involving $\Gamma^{\mu \nu}$ reads
\begin{eqnarray}
\nabla^{\mu} \Gamma^{\nu \lambda}+\nabla^{\nu} \Gamma^{\lambda \mu}+\nabla^{\lambda} \Gamma^{\mu \nu}=0.
\end{eqnarray}
An alternative form with building blocks is more useful,
\begin{align}
\nabla^\mu f_+^{\nu\lambda}+\nabla^\nu f_+^{\lambda\mu}+\nabla^\lambda f_+^{\mu\nu}
+\frac{i}{2}[u^{\mu},f_-^{\nu\lambda}]+\frac{i}{2}[u^{\nu},f_-^{\lambda\mu}]+\frac{i}{2}[u^{\lambda},f_-^{\mu\nu}]=0.
\end{align}

\subsubsection{Cayley-Hamilton relations}

All the building blocks in Sec. \ref{bbb} are $N_f\times N_f$ matrices in the flavor space. For any $2\times 2$ matrices $A$ and $B$, there exists a relation
\begin{eqnarray}
A B+B A-A\langle B\rangle-B\langle A\rangle-\langle A B\rangle+\langle A\rangle\langle B\rangle=0.
\end{eqnarray}
For any $3\times 3$ matrices $A$, $B$, and $C$, the relation is
\begin{align}
0={}& A B C+A C B+B A C+B C A+C A B+C B A \notag\\
&-A B\langle C\rangle-A C\langle B\rangle-B A\langle C\rangle-B C\langle A\rangle-C A\langle B\rangle \notag\\
&-C B\langle A\rangle-A\langle B C\rangle-B\langle A C\rangle-C\langle A B\rangle-\langle A B C\rangle \notag\\
&-\langle A C B\rangle+A\langle B\rangle\langle C\rangle+B\langle A\rangle\langle C\rangle+C\langle A\rangle\langle B\rangle \notag\\
&+\langle A\rangle\langle B C\rangle+\langle B\rangle\langle A C\rangle+\langle C\rangle\langle A B\rangle-\langle A\rangle\langle B\rangle\langle C\rangle.
\end{align}

\subsubsection{Contact terms}

For contact terms, the LR-basis is more convenient for the Lagrangian construction. From the building blocks in Eq. \eqref{buildingblocks}, one gets
\begin{align}
F_{L}^{\mu \nu} &=\frac{1}{2} u^{\dagger}\left(f_{+}^{\mu \nu}+f_{-}^{\mu \nu}\right) u,\\
F_{R}^{\mu \nu} &=\frac{1}{2} u\left(f_{+}^{\mu \nu}-f_{-}^{\mu \nu}\right) u^{\dagger}, \\
\chi &=\frac{1}{2} u\left(\chi_{+}+\chi_{-}\right) u, \\
\chi^{\dagger} &=\frac{1}{2} u^{\dagger}\left(\chi_{+}-\chi_{-}\right) u^{\dagger}.
\end{align}
Their transform properties can be found in Table \ref{tab:table1} of Ref. \cite{Fearing:1994ga}.

%%%%%%%%%%%%%%%%%%%%%%%%%%%%%%%%%%%%%%%%%%%%%%%%%%%%%%%%%%%%%%%%%%%%%%%%%%%%
\section{Relations between LECs in different cases}\label{sec:level1e}
%%%%%%%%%%%%%%%%%%%%%%%%%%%%%%%%%%%%%%%%%%%%%%%%%%%%%%%%%%%%%%%%%%%%%%%%%%%%

In the heavy diquark limit, the LECs in the relativistic Lagrangian are no longer independent. Some relations among these LECs appear. With these relations, the number of independent relativistic LECs could be reduced. A method to obtain these relations is to compare the relativistic Lagrangian with the Lagrangian in the heavy diquark limit. To remove the dimensions in the results and to simplify the forms of the relations, we define a new set of terms and LECs in the relativistic Lagrangians as follows,
\begin{eqnarray}
\tilde{O}_{n}=O_{n} / M^{r}, \quad \tilde{E}_{n}=E_{n} M^{r},
\end{eqnarray}
where $r$ is the number of covariant derivatives acting on the doubly charmed baryons. Now, all the $\tilde{E}_{n}$'s have the same dimension at a given order.

There are three types of Lorentz structures in the constructed terms of Eq. \eqref{eq:hlag}. With the definition of the superfield $\psi^{\mu}$ in Eq. \eqref{psi}, one gets
 \begin{align}
\bar{\psi}^{\mu}\psi_{\mu}\to {}&\bar{T}^{ \mu}T_{\mu}-\bar{B}B,\\
\bar{\psi}^{\mu}\gamma_{5}\gamma^{\lambda}\psi_{\mu}\to {}&\bar{T}^{ \mu}\gamma_{5}\gamma^{\lambda}T_{\mu}+\frac{1}{3}\bar{B}\gamma_{5}\gamma^{\lambda}B+\frac{2}{\sqrt{3}}\bar{T}^{ \lambda}B+\frac{2}{\sqrt{3}}\bar{B}{T}^{\lambda},\\
\bar{\psi}^{\mu}\sigma^{\lambda\rho}\psi_{\mu}\to {}&\bar{T}^{ \mu}\sigma^{\lambda\rho}T_{\mu}+\frac{1}{3}\overline{B}\sigma^{\lambda\rho}B
-\frac{2i}{\sqrt{3}}\bar{B}\gamma_{5}\gamma^{\rho}{T}^{ \lambda}\notag\\&+\frac{2i}{\sqrt{3}}\bar{B}\gamma_{5}\gamma^{\lambda}{T}^{ \rho}+\frac{2i}{\sqrt{3}}\bar{T}^{\lambda}\gamma_{5}\gamma^{\rho}{B}-\frac{2i}{\sqrt{3}}\bar{T}^{ \rho}\gamma_{5}\gamma^{\lambda}{B}.
\end{align}
The right-hand side expansions imply the possible LEC relations in the relativistic Lagrangians. Substituting the above structures into Eq. \eqref{eq:hlag} and comparing the relativistic and nonrelativistic terms, one can express the relations between $\tilde{E}_{m}$ and ${F}_{n}$ as
\begin{eqnarray}\label{LECrelations}
F_{n}= \sum_{m} \tilde{E}_{m} A_{m n}.
\end{eqnarray}
The number of ${F}_{n}$ ($N_F$) is less than the number of $\tilde{E}_{m}$ ($N_E$). Hence, these relations give $N_E-N_F$ constraint conditions among $\tilde{E}_{m}$ in the heavy diquark limit. With the help of linear algebra, one can pick up the independent relativistic LECs and eliminate the dependent ones. In this paper, we only consider the relations up to the order $\mathcal{O}(p^{3})$, which is restricted by computational conditions.

If one treats the heavy diquark as a compact object, the properties of the doubly charmed baryons can be related to those of the singly heavy mesons, which is the result of  the heavy diquark-antiquark symmetry \cite{Savage:1990di}. This symmetry then tells us that the doubly charmed baryons and heavy-light mesons with quark content $\bar{Q}q$ share the same LECs in ChPT \cite{Meng:2018zbl}. In Ref. \cite{Jiang:2019hgs}, we have obtained the chiral Lagrangian for heavy-light mesons with quark content $Q\bar{q}$ and the LEC relations in the heavy quark limit in that case. Noticing the charge-conjugation transformations for building blocks and superfields, one can understand that the LECs in $Q\bar{q}$ and $\bar{Q}q$ cases are the same. Now, the LEC relations between the doubly charmed baryon case and the heavy-light meson case could also be obtained by linear algebra.

\section{Results AND Discussions}\label{sec:level1f}

Now, we are ready to present the final results for the independent chiral-invariant terms. The details to get such terms are almost the same as those given in Refs. \cite{Jiang:2014via,Jiang:2016vax,Jiang:2017yda} and we do not talk about them any more. Here, the Lagrangians are expressed as
\begin{align}
&\mathcal{L}^{(m)}=\sum_{n} E_{n}^{(m)} O_{n}^{(m)}=\sum_{n} \tilde{E}_{n}^{(m)} \tilde{O}_{n}^{(m)},\quad N_{f}=3,\\
&\mathcal{L}^{(m)}=\sum_{n} e_{n}^{(m)} o_{n}^{(m)}=\sum_{n} \tilde{e}_{n}^{(m)} \tilde{o}_{n}^{(m)},\quad N_{f}=2,\\
&\mathcal{L}_{\mathrm{H}}^{(m)}=\sum_{n} F_{n}^{(m)} P_{n}^{(m)}\quad N_{f}=3,\\
&\mathcal{L}_{\mathrm{H}}^{(m)}=\sum_{n} f_{n}^{(m)} p_{n}^{(m)}\quad N_{f}=2,
\end{align}
where $m$ denotes the chiral dimension.

\subsection{$\mathcal{O}(p^{1})$ order}\label{sec:level2f1}

The relativistic result at the order $\mathcal{O}(p^{1})$ is \cite{Meng:2017dni}
\begin{align}
\mathcal{L}^{(1)}={}&\bar{B}(i\slashed{D}-m_{1})B+\bar{T}^{ \mu}[g_{\mu\nu}(i\slashed{D}-m_{2})+iA(\gamma_{\mu}D_{\nu}+\gamma_{\nu}D_{\mu})\notag\\
&+\frac{i}{2}(3A^{2}+2A+1)\gamma_{\mu}\slashed{D}\gamma_{\nu}+m_{2}(3A^{2}+3A+1)\gamma_{\mu}\gamma_{\nu}]T^{\nu}\notag\\
&+\frac{E_{1}}{2}\bar{B}u_{\mu}\gamma_{5}\gamma^{\mu}B+\frac{E_{2}}{2}\bar{T}^{ \mu}u_{\lambda}\gamma_{5}\gamma^{\lambda}T_{\mu}
+\frac{E_{3}}{2}(\bar{B}u^{\mu}T_{\mu}+\mathrm{H.c.}),
\end{align}
where $A\neq1/2$ is an unphysical parameter and it is usually taken to be $A=-1$. Generally speaking, the Lagrangian for spin-3/2 fields is invariant under the``point'' or ``contact'' transformation and the relations between different choices of $A$ can be found \cite{Jiang:2017yda}. Sine the value of $A$ does not affect the structure of chiral Lagrangians, we do not discuss more about it here. The Lagrangian at the leading order in the heavy diquark limit reads \cite{Meng:2018zbl}
\begin{align}
\mathcal{L}_{\mathrm{H} Q}^{(1)}=\bar{\psi}_\nu i v^{\mu} D_{\mu} \psi^\nu-\frac{1}{2} g\bar{\psi}_\nu u_{\lambda} \gamma_{5} \gamma^{\lambda} \psi^\nu.
\end{align}
For the relations between these LECs, one gets \cite{Meng:2018zbl}
\begin{align}\label{loworderrelations}
E_{1}=\frac{1}{3}E_{2}=\frac{1}{2\sqrt{3}}E_{3}=-\frac{1}{3}g.
\end{align}

\subsection{$\mathcal{O}(p^{2})$ order}

The interaction terms of $\mathcal{O}(p^{2})$ relativistic Lagrangian are collected in Table \ref{tab:table5}. The first column lists the forms of independent terms. The second and  sixth columns label the number for each term in the $SU(2)$ and $SU(3)$ cases, respectively. The terms without a number in the $SU(2)$ case mean that they are not independent. If one takes no account of the electromagnetic effect, the terms containing $\langle f_+^{\mu\nu}\rangle$ could be ignored. The third and the seventh columns show the relations between the relativistic LECs and the LECs in the heavy diquark limit. To assign the independent terms in the relativistic Lagrangian, we put a symbol ``I'' in the forth and eighth columns. These two columns also present the relations of the dependent LECs to the independent ones. The fifth and ninth columns give the LEC relations in the heavy diquark-antiquark symmetry, where $\tilde{c}$ and $\tilde{C}$ defined in Ref. \cite{Jiang:2019hgs} are LECs for heavy-light mesons. We show the $\mathcal{O}(p^{2})$ Lagrangian terms for the superfield $\psi^\mu$ in Table \ref{tab:table4}.

From Table \ref{tab:table5}, the numbers of $\mathcal{O}(p^{2})$ terms in ${\cal L}_{BB}$, ${\cal L}_{TT}$, and ${\cal L}_{BT}$ in the $SU(2)$ case are 7, 8, and 4, respectively. Those in the $SU(3)$ case are 9, 11, and 5, respectively. From Table \ref{tab:table4}, the heavy diquark symmetry reduces significantly the total number of independent terms from 19 (25) to be 7 (8) in the $SU(2)$ ($SU(3)$) case.

{\renewcommand\arraystretch{1.3}
\begin{longtable}{lcrrrcrrr}
\caption{Independent terms in the relativistic Lagrangian at the order $\mathcal{O}(p^{2})$. The columns 2, 3, 4, and 5 (6, 7, 8, and 9) are for the two (three)-flavor case. The columns 2 and 6 label the number for each term. The terms without a number are not independent. The columns 3 and 7 list the relations between LECs in the relativistic case and those in the heavy diquark limit. The columns 4 and 8 show the LEC relations among different terms in the relativistic case by using the heavy diquark symmetry. ``I'' means that the term is chosen as an independent term in the heavy diquark limit. The columns 5 and 9 give the LEC relations with the help of the heavy diquark-antiquark symmetry, where $\tilde{c}$ and $\tilde{C}$ are LECs in the heavy-light meson case. Such LECs are defined in Ref. \cite{Jiang:2019hgs}. ``0'' means that the LECs in the heavy diquark limit vanish. }\label{tab:table5}
\\
\hline\hline $O_n/o_n$ & $SU(2)$ & $\tilde{e}^{(2)}_n$ &$\tilde{e}^{(2)}_n$ & $\tilde{e}^{(2)}_n$ & $SU(3)$ & $\tilde{E}^{(2)}_n$&$\tilde{E}^{(2)}_n$ & $\tilde{E}^{(3)}_n$\\
\hline\endfirsthead

\hline\hline $O_n$ & $SU(2)$ & $\tilde{e}^{(2)}_n$ &$\tilde{c}^{(2)}_n$ & $\tilde{e}^{(2)}_n$ & $SU(3)$ & $\tilde{E}^{(2)}_n$&$\tilde{C}^{(2)}_n$ & $\tilde{E}^{(3)}_n$\\
\hline\endhead

\hline\hline % \multicolumn{8}{c}{Continued on next page}\\
\endfoot

\hline\endlastfoot

$\bar{B}\left\langle u^{\mu}u_{\mu}\right\rangle B$&1&$-f_{1}^{(2)}$&I&$\frac{1}{4}\tilde{c}_{1}^{(2)}$&1&$-F_{1}^{(2)}$&I&$\frac{1}{2}\tilde{C}_{3}^{(2)}$\\$\bar{B}\left\langle u^{\mu}u^{\nu}\right\rangle D_{\mu\nu}B$&2&$f_{2}^{(2)}$&I&$\frac{1}{4}\tilde{c}_{2}^{(2)}$&2&$F_{2}^{(2)}$&I&$\frac{1}{2}\tilde{C}_{4}^{(2)}$\\$\bar{B}u^{\mu}u_{\mu}B$&&&&&3&$-F_{3}^{(2)}$&I&$\frac{1}{2}\tilde{C}_{1}^{(2)}$\\$\bar{B}u^{\mu}u^{\nu}D_{\mu\nu}B$&&&&&4&$F_{4}^{(2)}$&I&$\frac{1}{2}\tilde{C}_{2}^{(2)}$\\$i\bar{B}u^{\mu}u^{\nu}\sigma_{\mu\nu}B$&3&$\frac{1}{3}f_{3}^{(2)}$&I&$-\frac{1}{6}\tilde{c}_{5}^{(2)}$&5&$\frac{1}{3}F_{5}^{(2)}$&I&$-\frac{1}{6}\tilde{C}_{7}^{(2)}$\\$\bar{B}\left\langle {f_{+}}^{\mu\nu}\right\rangle \sigma_{\mu\nu}B$&4&$\frac{1}{3}f_{4}^{(2)}$&I&$\frac{1}{12}\tilde{c}_{10}^{(2)}$&6&$0$&$0$&$0$\\$\bar{B}{f_{+}}^{\mu\nu}\sigma_{\mu\nu}B$&5&$\frac{1}{3}f_{5}^{(2)}$&I&$\frac{1}{12}\tilde{c}_{9}^{(2)}$&7&$\frac{1}{3}F_{6}^{(2)}$&I&$\frac{1}{12}\tilde{C}_{14}^{(2)}$\\$\bar{B}\left\langle \chi_{+}\right\rangle B$&6&$-f_{6}^{(2)}$&I&$\frac{1}{2}\tilde{c}_{4}^{(2)}$&8&$-F_{7}^{(2)}$&I&$\frac{1}{2}\tilde{C}_{6}^{(2)}$\\$\bar{B}\chi_{+}B$&7&$-f_{7}^{(2)}$&I&$\frac{1}{2}\tilde{c}_{3}^{(2)}$&9&$-F_{8}^{(2)}$&I&$\frac{1}{2}\tilde{C}_{5}^{(2)}$\\$\bar{T}^{\mu}\left\langle u_{\mu}u^{\nu}\right\rangle T_{\nu}$&8&$f_{3}^{(2)}$&$3\tilde{e}_{3}^{(2)}$&$-\frac{1}{2}\tilde{c}_{5}^{(2)}$&10&$0$&$0$&$0$\\$\bar{T}^{\mu}\left\langle u^{\nu}u_{\nu}\right\rangle T_{\mu}$&9&$f_{1}^{(2)}$&$-\tilde{e}_{1}^{(2)}$&$-\frac{1}{4}\tilde{c}_{1}^{(2)}$&11&$F_{1}^{(2)}$&$-\tilde{E}_{1}^{(2)}$&$-\frac{1}{2}\tilde{C}_{3}^{(2)}$\\$\bar{T}^{\mu}\left\langle u^{\nu}u^{\lambda}\right\rangle D_{\nu\lambda}T_{\mu}$&10&$-f_{2}^{(2)}$&$-\tilde{e}_{2}^{(2)}$&$-\frac{1}{4}\tilde{c}_{2}^{(2)}$&12&$-F_{2}^{(2)}$&$-\tilde{E}_{2}^{(2)}$&$-\frac{1}{2}\tilde{C}_{4}^{(2)}$\\$\bar{T}^{\mu}u_{\mu}u^{\nu}T_{\nu}$&11&$-2f_{3}^{(2)}$&$-6\tilde{e}_{3}^{(2)}$&$\tilde{c}_{5}^{(2)}$&13&$-F_{5}^{(2)}$&$-3\tilde{E}_{5}^{(2)}$&$\frac{1}{2}\tilde{C}_{7}^{(2)}$\\$\bar{T}^{\mu}u^{\nu}u_{\mu}T_{\nu}$&&&&&14&$F_{5}^{(2)}$&$3\tilde{E}_{5}^{(2)}$&$-\frac{1}{2}\tilde{C}_{7}^{(2)}$\\$\bar{T}^{\mu}u^{\nu}u_{\nu}T_{\mu}$&&&&&15&$F_{3}^{(2)}$&$-\tilde{E}_{3}^{(2)}$&$-\frac{1}{2}\tilde{C}_{1}^{(2)}$\\$\bar{T}^{\mu}u^{\nu}u^{\lambda}D_{\nu\lambda}T_{\mu}$&&&&&16&$-F_{4}^{(2)}$&$-\tilde{E}_{4}^{(2)}$&$-\frac{1}{2}\tilde{C}_{2}^{(2)}$\\$i\bar{T}^{\mu}\left\langle {f_{+\mu}}^{\nu}\right\rangle T_{\nu}$&12&$2f_{4}^{(2)}$&$6\tilde{e}_{4}^{(2)}$&$\frac{1}{2}\tilde{c}_{10}^{(2)}$&17&$0$&$0$&$0$\\$i\bar{T}^{\mu}{f_{+\mu}}^{\nu}T_{\nu}$&13&$2f_{5}^{(2)}$&$6\tilde{e}_{5}^{(2)}$&$\frac{1}{2}\tilde{c}_{9}^{(2)}$&18&$2F_{6}^{(2)}$&$6\tilde{E}_{6}^{(2)}$&$\frac{1}{2}\tilde{C}_{14}^{(2)}$\\$\bar{T}^{\mu}\left\langle \chi_{+}\right\rangle T_{\mu}$&14&$f_{6}^{(2)}$&$-\tilde{e}_{6}^{(2)}$&$-\frac{1}{2}\tilde{c}_{4}^{(2)}$&19&$F_{7}^{(2)}$&$-\tilde{E}_{7}^{(2)}$&$-\frac{1}{2}\tilde{C}_{6}^{(2)}$\\$\bar{T}^{\mu}\chi_{+}T_{\mu}$&15&$f_{7}^{(2)}$&$-\tilde{e}_{7}^{(2)}$&$-\frac{1}{2}\tilde{c}_{3}^{(2)}$&20&$F_{8}^{(2)}$&$-\tilde{E}_{8}^{(2)}$&$-\frac{1}{2}\tilde{C}_{5}^{(2)}$\\$\bar{B}\left\langle u^{\mu}u^{\nu}\right\rangle \gamma_{5}\gamma_{\mu}T_{\nu}+\mathrm{H.c.}$&16&$\frac{2}{\sqrt{3}}f_{3}^{(2)}$&$2\sqrt{3}\tilde{e}_{3}^{(2)}$&$-\frac{1}{\sqrt{3}}\tilde{c}_{5}^{(2)}$&21&$0$&$0$&$0$\\$\bar{B}u^{\mu}u^{\nu}\gamma_{5}\gamma_{\mu}T_{\nu}+\mathrm{H.c.}$&17&$-\frac{4}{\sqrt{3}}f_{3}^{(2)}$&$-4\sqrt{3}\tilde{e}_{3}^{(2)}$&$\frac{2}{\sqrt{3}}\tilde{c}_{5}^{(2)}$&22&$-\frac{2}{\sqrt{3}}F_{5}^{(2)}$&$-2\sqrt{3}\tilde{E}_{5}^{(2)}$&$\frac{1}{\sqrt{3}}\tilde{C}_{7}^{(2)}$\\$\bar{B}u^{\mu}u^{\nu}\gamma_{5}\gamma_{\nu}T_{\mu}+\mathrm{H.c.}$&&&&&23&$\frac{2}{\sqrt{3}}F_{5}^{(2)}$&$2\sqrt{3}\tilde{E}_{5}^{(2)}$&$-\frac{1}{\sqrt{3}}\tilde{C}_{7}^{(2)}$\\$i\bar{B}\left\langle {f_{+}}^{\mu\nu}\right\rangle \gamma_{5}\gamma_{\mu}T_{\nu}+\mathrm{H.c.}$&18&$\frac{4}{\sqrt{3}}f_{4}^{(2)}$&$4\sqrt{3}\tilde{e}_{4}^{(2)}$&$\frac{1}{\sqrt{3}}\tilde{c}_{10}^{(2)}$&24&$0$&$0$&$0$\\$i\bar{B}{f_{+}}^{\mu\nu}\gamma_{5}\gamma_{\mu}T_{\nu}+\mathrm{H.c.}$&19&$\frac{4}{\sqrt{3}}f_{5}^{(2)}$&$4\sqrt{3}\tilde{e}_{5}^{(2)}$&$\frac{1}{\sqrt{3}}\tilde{c}_{9}^{(2)}$&25&$\frac{4}{\sqrt{3}}F_{6}^{(2)}$&$4\sqrt{3}\tilde{E}_{6}^{(2)}$&$\frac{1}{\sqrt{3}}\tilde{C}_{14}^{(2)}$\\
\hline
\end{longtable}}

\begin{table}[htbp]
\caption{Independent Lagrangian terms in the heavy diquark limit at the order $\mathcal{O}(p^{2})$. The second (third) column labels the number for each term in the two (three)-flavor case. The terms without a number are not independent.}\label{tab:table4}
\begin{ruledtabular}
\begin{tabular}{lcc}
$P_n/p_n$  & $SU(2)$ & $SU(3)$\\\hline
$\psib^{\mu}\la u^{\nu}u_{\nu}\ra\psi_{\mu}$ & 1 & 1  \\
$\psib^{\mu}\la u^{\nu}u^{\lambda}\ra v_{\nu}v_{\lambda}\psi_{\mu}$ & 2 & 2  \\
$\psib^{\mu}u^{\nu}u_{\nu}\psi_{\mu}$ &  & 3  \\
$\psib^{\mu}u^{\nu}u^{\lambda}v_{\nu}v_{\lambda}\psi_{\mu}$ &  & 4  \\
$i\psib^{\mu}u^{\nu}u^{\lambda}\sigma_{\nu\lambda}\psi_{\mu}$ & 3 & 5  \\
$\psib^{\mu}\la{f_{+}}^{\nu\lambda}\ra\sigma_{\nu\lambda}\psi_{\mu}$ & 4 &  \\
$\psib^{\mu}{f_{+}}^{\nu\lambda}\sigma_{\nu\lambda}\psi_{\mu}$ & 5 & 6  \\
$\psib^{\mu}\la\chip\ra\psi_{\mu}$ & 6 & 7  \\
$\psib^{\mu}\chip\psi_{\mu}$ & 7 & 8  \\
\end{tabular}
\end{ruledtabular}
\end{table}

\subsection{$\mathcal{O}(p^{3})$ and $\mathcal{O}(p^{4})$ orders}

The numbers of interaction terms are large at the $\mathcal{O}(p^{3})$ and $\mathcal{O}(p^{4})$ orders. We list the third and fourth order results in Appendix \ref{p3r} and \ref{p4r}, respectively. Table \ref{tab:table7} (\ref{tab:table6}) in Appendix \ref{p3r} is for the $\mathcal{O}(p^{3})$ relativistic (nonrelativistic) terms. Since the LEC relations may be long, part of them are given outside the Table \ref{tab:table7}. The $\mathcal{O}(p^{4})$ relativistic (nonrelativistic) Lagrangian terms are shown in Table \ref{tab:table9} (\ref{tab:table8}) in Appendix \ref{p4r}. In Table \ref{tab:table9}, we only label the number for each term and mark the independent terms in the heavy diquark limit.

From Table \ref{tab:table7}, the numbers of $\mathcal{O}(p^{3})$ terms in ${\cal L}_{BB}$, ${\cal L}_{TT}$, and ${\cal L}_{BT}$ in the $SU(2)$ case are 23, 26, and 25, respectively. Those in the $SU(3)$ case are 33, 38, and 41, respectively. From Table \ref{tab:table6}, the heavy diquark symmetry reduces significantly the total number of independent terms from 74 (112) to be 23 (31) in the $SU(2)$ ($SU(3)$) case.

From Table \ref{tab:table9}, the numbers of $\mathcal{O}(p^{4})$ terms in ${\cal L}_{BB}$, ${\cal L}_{TT}$, and ${\cal L}_{BT}$ in the $SU(2)$ case are 118, 154, and 180, respectively. Those in the $SU(3)$ case are 216, 304, and 344, respectively. From Table \ref{tab:table8}, the heavy diquark symmetry reduces significantly the total number of independent terms from 452 (864) to be 118 (189) in the $SU(2)$ ($SU(3)$) case.

\subsection{Discussions}\label{sec:level2f2}

In the literature, one may find part of the constructed chiral Lagrangians for the doubly charmed baryons. Here, we make some comparisons. Our $\mathcal{L}_{BB}$ in the two-flavor case has the same number of independent terms as that in Ref. \cite{Fettes:2000gb}, but the choice of independent structures is different. In the three-flavor case, the terms in our $\mathcal{L}_{BB}$ are consistent with those in Ref. \cite{Qiu:2020omj} at the first two orders, but we have one less term at the third chiral order. In fact, the 32nd term $O^{(3)}_{32}=\bar{\psi}\left[D^{\lambda}, f_{-}^{\mu \nu}\right] \gamma_{5} \gamma_{\mu} D_{\nu \lambda}\psi+\mathrm {H.c.}$ over there can be eliminated with Eq. (2.41) of Ref. \cite{Fettes:2000gb}. At the fourth order, there exist some redundant terms similarly, such as $O^{(4)}_{195}=\bar{\psi}[D^{\mu\nu}, f_{+}^{\lambda\rho}]\sigma_{\lambda\rho}D_{\mu\nu}\psi+\mathrm {H.c.}$, $O^{(4)}_{196}=\bar{\psi}[D^{\mu\nu}, f_{+}^{\lambda\rho}]\sigma_{\mu\lambda}D_{\nu\rho}\psi+\mathrm {H.c.}$, and so on. Moreover, linear relations, such as Cayley-Hamilton relations, are not completely used. We have checked some of these new relations by hand explicitly and have removed all redundant terms by computer. Hence, we have much fewer terms at the $\mathcal{O}(p^{4})$ order. In the heavy diquark limit, our results are consistent with the $\mathcal{O}(p^{2})$ terms obtained in Ref. \cite{Meng:2018zbl}.

One may wonder whether the ${\cal O}(p^4)$ order Lagrangian is useful because the number of LECs is very large. At present, there are studies of masses \cite{Yao:2018ifh,Sun:2014aya} and magnetic moments \cite{Meng:2017dni,Li:2017pxa} for the doubly heavy baryons with chiral Lagrangians up to the ${\cal O}(p^4)$ order. The complete Lagrangian we obtain can be used to confirm the terms constructed in the literature. A calculated observable contains different order LECs and one needs to determine the involved LECs by fitting available data. Usually, several LECs or LEC combinations are involved in the study of a special problem. One may, in principle, use various theoretical methods (quark model, large $N_c$, lattice QCD, etc) to determine or constrain such LEC values. It is true that available data to determine the ${\cal O}(p^4)$ LECs in the existing studies about doubly heavy baryons are still scarce. However, once the LECs are determined, the convergence of chiral expansion, an important issue one should check in understanding the power counting problem in ChPT, can be certainly better understood in the investigations up to the ${\cal O}(p^4)$ order than that to the ${\cal O}(p^3)$ order. Moreover, the complete Lagrangian can serve for future studies of any other quantities.

In fact, the complete Lagrangian may motivate future studies of LEC determinations or LEC relations. Since the numbers of LECs are significantly increased for high-order terms, theoretical methods to study LECs in a systematic way are welcome. In Ref. \cite{Jiang:2015dba}, an analytical method was adopted to calculate the LECs for mesonic chiral Lagrangians. A possible method in the present case would also be developed, which needs complete Lagrangians with independent terms. In Refs. \cite{Jiang:2022gjy,Jiang:2023zqq}, we preliminarily studied the LEC relations with the help of the chiral quark model up to the third chiral order in the light baryon cases, which is based on operator correspondences between hadron-level structures and quark-level structures. In principle, such a study can be extended to any chiral order, but it also needs complete Lagrangians with independent terms.

Since the LEC determination is very important in the application of ChPT, here we discuss a little more about the chiral quark model method which is feasible in constraining LECs before enough data are available. In Ref. \cite{Jiang:2022gjy}, we found LEC relations between the $SU(2)$ $\pi N$ Lagrangians and the $SU(3)$ meson-baryon Lagrangians, by employing the idea that the hadron-level interactions can be equivalently described in the chiral quark model and by noticing that quarks are placed in the fundamental representations of both $SU(2)$ and $SU(3)$. Since the LECs in the $\pi N$ Lagrangians can be extracted from a large number of experimental data, one may get some constraints on LECs in the $SU(3)$ $\phi B$ Lagrangians with $\phi$ ($B$) being the pseudoscalar meson (octet baryon). Similar to the Lagrangian construction at the hadron level, we have constructed high-order Lagrangians at the quark level. Such Lagrangians may be applied to all the $qqq$, $Qqq$, and $QQq$ baryon cases where $q$ ($Q$) represents a light (heavy) quark. In Ref. \cite{Jiang:2023zqq}, when studying the LEC relations for the spin-3/2 $\Delta$ baryon case, we found that several multiplication factors to be determined are actually needed. At that time, we only determined values of two such factors from a phenomenological perspective. It is possible to determine such factors with the heavy quark symmetry by using the chiral Lagrangians with $QQq$ and $Qqq$ \cite{Zou:2023uhr} baryons. Since only one light quark exists in the doubly charmed baryon and the pion interaction is mainly induced by the light quark, it is obvious that the LECs may have some relations with the coupling parameters in the chiral quark model. In the heavy diquark limit, the heavy quarks decouple with the light quark and the spin-1/2 and spin-3/2 doubly charmed baryons can be described simultaneously by the superfield $\psi^\mu$. The number of independent Lagrangian terms for $\psi^\mu$ should be the same as that in the chiral quark model description. One may confirm this by consulting Ref. \cite{Jiang:2022gjy} for the relevant Lagrangian up to the third chiral order. Since we mainly focus on the search for a minimal set of Lagrangian terms in the present paper, we leave the discussions to get LEC relations and to determine the multiplication factors for spin-3/2 baryons in a separate work.

It is usually interesting to know how many independent structures there are for the Lagrangian at a certain order in ChPT. The constructed Lagrangian with explicit terms can surely give an answer. Besides the method to construct Lagrangians, it is also possible to find an answer with other approaches. In the mesonic case \cite{Graf:2020yxt}, the Hilbert series techniques have been used to analyze numbers of independent terms at different orders, where the role of the concrete Lagrangians is to validate the method. When a similar study for the doubly heavy baryon case is performed, the Lagrangian constructed here would also be helpful to validate the analysis.

If one replaces the $cc$ diquark by the $bb$ diquark in the doubly charmed baryons, the chiral Lagrangians for the doubly bottom baryons are obtained. For the $bcq$ baryons, the situation is slightly different. In the case that the spin of the $bc$ diquark is 1, the chiral Lagrangians have the same structures discussed above. Since the spin of the $bc$ diqaurk can also be 0, an additional flavor triplet exists for the spin-1/2 $bcq$ baryons. The Lagrangian structures for the spin-1/2 $bcq$ baryons in the scalar and axial-vector $bc$ diquark cases are the same.

%%%%%%%%%%%%%%%%%%%%%%%%%%%%%%%%%%%%%%%%
\section{Summary}\label{sec:level1g}
%%%%%%%%%%%%%%%%%%%%%%%%%%%%%%%%%%%%%%%%

In this paper, we constructed the relativistic chiral Lagrangians for both spin-$\frac{1}{2}$ and spin-$\frac{3}{2}$ doubly charmed baryons up to the order $\mathcal{O}(p^{4})$, in both two- and three-flavor cases. The chiral Lagrangians in the heavy diquark limit are also obtained up to the order $\mathcal{O}(p^{4})$. Table \ref{numberofterms} collects the number of independent terms at each order. It seems that the numbers in the heavy diquark limit are about one third (one forth) of those in the relativistic case at the order $\mathcal{O}(p^{3})$ ($\mathcal{O}(p^{4})$). Obviously, the heavy diquark symmetry is helpful for us to reduce the number of unknown parameters in employing ChPT. We present the LEC relations between the relativistic case and the case in the heavy diquark limit up to the order $\mathcal{O}(p^{3})$. In addition, the heavy diquark-antiquark symmetry is also considered. With this symmetry, we obtain the LEC relations between the doubly charmed baryon case and the heavy-light meson case.

\begin{table}[htbp]
\caption{Number of independent terms at each chiral order.}\label{numberofterms}
\begin{tabular}{ccccc|ccccc}\hline\hline
&\multicolumn{4}{c}{Relativistic Lagrangian}&\multicolumn{4}{c}{Nonrelativistic Lagrangian}\\
Chiral order&${\cal O}(p^1)$&${\cal O}(p^2)$&${\cal O}(p^3)$&${\cal O}(p^4)$&${\cal O}(p^1)$&${\cal O}(p^2)$&${\cal O}(p^3)$&${\cal O}(p^4)$\\
\hline$SU(2)$&3&19&74&452&1&7&23&118\\
$SU(3)$&3&25&112&864&1&8&31&189\\\hline\hline
\end{tabular}
\end{table}

\begin{acknowledgments}
This project was supported by the National Natural Science Foundation of China under Grants No. 11775132, No. 12235008, and Guangxi Science Foundation under Grants No. 2022GXNSFAA035489.
\end{acknowledgments}

\appendix
\section{$\mathcal{O}(p^{3})$ order results}\label{p3r}
This appendix gives Tables \ref{tab:table7} and \ref{tab:table6}.
\begin{longtable}{lcrrrcrrr}
\caption{Independent terms in the relativistic Lagrangian at the order $\mathcal{O}(p^{3})$. The columns 2, 3, 4, and 5 (6, 7, 8, and 9) are for the two (three)-flavor case. The columns 2 and 6 label the number for each term. The terms without a number are not independent in the $SU(2)$ case. The columns 3 and 7 list the relations between LECs in the relativistic case and those in the heavy diquark limit. The columns 4 and 8 show the LEC relations among different terms in the relativistic case by using the heavy diquark symmetry. ``I'' means that the term is chosen as an independent term in the heavy diquark limit. The columns 5 and 9 give the LEC relations with the help of the heavy diquark-antiquark symmetry, where $\tilde{c}$ and $\tilde{C}$ are LECs in the heavy-light meson case. Such LECs are defined in Ref. \cite{Jiang:2019hgs}. ``0'' means that the LECs in the heavy diquark limit vanish. The symbol ``*'' indicates that the relation is long and we show relevant results in Eqs. \eqref{eqA1}-\eqref{eqA6} in this appendix.}\label{tab:table7}
\\
\hline\hline $O_n/o_n$ & $SU(2)$ & $\tilde{e}^{(3)}_n$  & $\tilde{e}^{(3)}_n$&$\tilde{e}^{(3)}_n$ & $SU(3)$ & $\tilde{E}^{(3)}_n$ & $\tilde{E}^{(3)}_n$&$\tilde{E}^{(3)}_n$\\
\hline\endfirsthead

\hline\hline $O_n/o_n$ & $SU(2)$ & $\tilde{e}^{(3)}_n$  & $\tilde{e}^{(3)}_n$&$\tilde{e}^{(3)}_n$ & $SU(3)$ & $\tilde{E}^{(3)}_n$ & $\tilde{E}^{(3)}_n$&$\tilde{E}^{(3)}_n$\\
\hline\endhead

\hline\hline % \multicolumn{8}{c}{Continued on next page}\\
\endfoot

\hline\endlastfoot
$\bar{B}\left\langle u^{\mu}u_{\mu}\right\rangle u^{\nu}\gamma_{5}\gamma_{\nu}B$&1&$\frac{1}{3}f_{1}^{(3)}$&I&$*$&1&$\frac{1}{3}F_{1}^{(3)}$&I&$*$\\$\bar{B}\left\langle u^{\mu}u^{\nu}\right\rangle u_{\mu}\gamma_{5}\gamma_{\nu}B$&2&$\frac{1}{3}f_{2}^{(3)}$&I&$-\frac{1}{12}\tilde{c}_{10}^{(3)}$&2&$\frac{1}{3}F_{2}^{(3)}$&I&$*$\\$\bar{B}\left\langle u^{\mu}u^{\nu}\right\rangle u^{\lambda}\gamma_{5}\gamma_{\mu}D_{\nu\lambda}B$&3&$-\frac{1}{3}f_{3}^{(3)}$&I&$*$&3&$-\frac{1}{3}F_{3}^{(3)}$&I&$*$\\$\bar{B}\left\langle u^{\mu}u^{\nu}\right\rangle u^{\lambda}\gamma_{5}\gamma_{\lambda}D_{\mu\nu}B$&4&$-\frac{1}{3}f_{4}^{(3)}$&I&$*$&4&$-\frac{1}{3}F_{4}^{(3)}$&I&$*$\\$\bar{B}\left\langle u^{\mu}u_{\mu}u^{\nu}\right\rangle \gamma_{5}\gamma_{\nu}B$&&&&&5&$\frac{1}{3}F_{5}^{(3)}$&I&$*$\\$\bar{B}\left\langle u^{\mu}u^{\nu}u^{\lambda}\right\rangle \gamma_{5}\gamma_{\mu}D_{\nu\lambda}B$&&&&&6&$-\frac{1}{3}F_{6}^{(3)}$&I&$*$\\$\bar{B}u^{\mu}u_{\mu}u^{\nu}\gamma_{5}\gamma_{\nu}B+\mathrm{H.c.}$&&&&&7&$\frac{1}{3}F_{7}^{(3)}$&I&$*$\\$\bar{B}u^{\mu}u^{\nu}u^{\lambda}\gamma_{5}\gamma_{\mu}D_{\nu\lambda}B+\mathrm{H.c.}$&&&&&8&$-\frac{1}{3}F_{8}^{(3)}$&I&$*$\\$\epsilon^{\mu\nu\lambda\rho}\bar{B}\left\langle u_{\mu}u_{\nu}u_{\lambda}\right\rangle D_{\rho}B$&5&$f_{5}^{(3)}$&I&$-\tilde{c}_{1}^{(3)}$&9&$F_{9}^{(3)}$&I&$-\frac{1}{2}\tilde{C}_{2}^{(3)}$\\$\epsilon^{\mu\nu\lambda\rho}\bar{B}u_{\mu}u_{\nu}u_{\lambda}D_{\rho}B$&&&&&10&$F_{10}^{(3)}$&I&$-\frac{1}{2}\tilde{C}_{1}^{(3)}$\\$i\bar{B}\left\langle u^{\mu}{f_{-}}^{\nu\lambda}\right\rangle \sigma_{\mu\nu}D_{\lambda}B$&6&$\frac{1}{3}f_{13}^{(3)}$&I&$-\frac{1}{6}\tilde{c}_{15}^{(3)}$&11&$\frac{1}{3}F_{19}^{(3)}$&I&$-\frac{1}{6}\tilde{C}_{30}^{(3)}$\\$i\bar{B}\left\langle u^{\mu}{f_{-}}^{\nu\lambda}\right\rangle \sigma_{\nu\lambda}D_{\mu}B$&7&$\frac{1}{3}f_{14}^{(3)}$&I&$\frac{1}{12}\tilde{c}_{16}^{(3)}$&12&$\frac{1}{3}F_{20}^{(3)}$&I&$\frac{1}{12}\tilde{C}_{31}^{(3)}$\\$i\bar{B}\left\langle u^{\mu}h^{\nu\lambda}\right\rangle \sigma_{\mu\nu}D_{\lambda}B$&8&$\frac{1}{3}f_{6}^{(3)}$&I&$\frac{1}{6}\tilde{c}_{19}^{(3)}$&13&$\frac{1}{3}F_{11}^{(3)}$&I&$\frac{1}{6}\tilde{C}_{32}^{(3)}$\\$\bar{B}u^{\mu}{f_{-\mu}}^{\nu}D_{\nu}B+\mathrm{H.c.}$&9&$f_{15}^{(3)}$&I&$-\frac{1}{2}\tilde{c}_{2}^{(3)}$&14&$F_{21}^{(3)}$&I&$-\frac{1}{2}\tilde{C}_{3}^{(3)}$\\$i\bar{B}u^{\mu}{f_{-}}^{\nu\lambda}\sigma_{\mu\nu}D_{\lambda}B+\mathrm{H.c.}$&&&&&15&$\frac{1}{3}F_{22}^{(3)}$&I&$-\frac{1}{6}\tilde{C}_{21}^{(3)}$\\$i\bar{B}u^{\mu}{f_{-}}^{\nu\lambda}\sigma_{\nu\lambda}D_{\mu}B+\mathrm{H.c.}$&&&&&16&$\frac{1}{3}F_{23}^{(3)}$&I&$\frac{1}{12}\tilde{C}_{22}^{(3)}$\\$\bar{B}u^{\mu}{h_{\mu}}^{\nu}D_{\nu}B+\mathrm{H.c.}$&10&$f_{9}^{(3)}$&I&$-\frac{1}{2}\tilde{c}_{3}^{(3)}$&17&$F_{14}^{(3)}$&I&$-\frac{1}{2}\tilde{C}_{4}^{(3)}$\\$\bar{B}u^{\mu}h^{\nu\lambda}D_{\mu\nu\lambda}B+\mathrm{H.c.}$&11&$-f_{10}^{(3)}$&I&$-\frac{1}{2}\tilde{c}_{4}^{(3)}$&18&$-F_{15}^{(3)}$&I&$-\frac{1}{2}\tilde{C}_{5}^{(3)}$\\$i\bar{B}u^{\mu}h^{\nu\lambda}\sigma_{\mu\nu}D_{\lambda}B+\mathrm{H.c.}$&&&&&19&$\frac{1}{3}F_{16}^{(3)}$&I&$\frac{1}{6}\tilde{C}_{25}^{(3)}$\\$\bar{B}\nabla^{\mu}{f_{-\mu}}^{\nu}\gamma_{5}\gamma_{\nu}B$&12&$\frac{1}{3}f_{18}^{(3)}$&I&$\frac{1}{6}\tilde{c}_{24}^{(3)}$&20&$\frac{1}{3}F_{25}^{(3)}$&I&$\frac{1}{6}\tilde{C}_{33}^{(3)}$\\$i\bar{B}{f_{+}}^{\mu\nu}u_{\mu}\gamma_{5}\gamma_{\nu}B+\mathrm{H.c.}$&13&$\frac{1}{3}f_{7}^{(3)}$&I&$\frac{1}{6}\tilde{c}_{30}^{(3)}$&21&$\frac{1}{3}F_{12}^{(3)}$&I&$\frac{1}{6}\tilde{C}_{39}^{(3)}$\\$i\bar{B}{f_{+}}^{\mu\nu}u^{\lambda}\gamma_{5}\gamma_{\mu}D_{\nu\lambda}B+\mathrm{H.c.}$&14&$-\frac{1}{3}f_{8}^{(3)}$&I&$\frac{1}{6}\tilde{c}_{41}^{(3)}$&22&$-\frac{1}{3}F_{13}^{(3)}$&I&$\frac{1}{3}\tilde{C}_{37}^{(3)}$\\$i\epsilon^{\mu\nu\lambda\rho}\bar{B}\left\langle f_{+\mu\nu}\right\rangle u_{\lambda}D_{\rho}B$&15&$-f_{11}^{(3)}$&I&$-\frac{1}{2}\tilde{c}_{6}^{(3)}$&23&$0$&$0$&$0$\\$i\epsilon^{\mu\nu\lambda\rho}\bar{B}\left\langle f_{+\mu\nu}u_{\lambda}\right\rangle D_{\rho}B$&16&$-f_{12}^{(3)}$&I&$*$&24&$-F_{17}^{(3)}$&I&$-\frac{1}{2}\tilde{C}_{7}^{(3)}$\\$i\epsilon^{\mu\nu\lambda\rho}\bar{B}f_{+\mu\nu}u_{\lambda}D_{\rho}B+\mathrm{H.c.}$&&&&&25&$-F_{18}^{(3)}$&I&$-\frac{1}{2}\tilde{C}_{6}^{(3)}$\\$i\bar{B}\left\langle \nabla^{\mu}{f_{+\mu}}^{\nu}\right\rangle D_{\nu}B$&17&$-f_{16}^{(3)}$&I&$-\frac{1}{2}\tilde{c}_{8}^{(3)}$&26&$0$&$0$&$0$\\$i\bar{B}\nabla^{\mu}{f_{+\mu}}^{\nu}D_{\nu}B$&18&$-f_{17}^{(3)}$&I&$-\frac{1}{2}\tilde{c}_{7}^{(3)}$&27&$-F_{24}^{(3)}$&I&$-\frac{1}{2}\tilde{C}_{8}^{(3)}$\\$\bar{B}\left\langle u^{\mu}\chi_{+}\right\rangle \gamma_{5}\gamma_{\mu}B$&19&$\frac{1}{3}f_{19}^{(3)}$&I&$-\frac{1}{6}\tilde{c}_{38}^{(3)}$&28&$\frac{1}{3}F_{26}^{(3)}$&I&$-\frac{1}{6}\tilde{C}_{45}^{(3)}$\\$\bar{B}\left\langle \chi_{+}\right\rangle u^{\mu}\gamma_{5}\gamma_{\mu}B$&20&$\frac{1}{3}f_{20}^{(3)}$&I&$*$&29&$\frac{1}{3}F_{27}^{(3)}$&I&$-\frac{1}{6}\tilde{C}_{46}^{(3)}$\\$\bar{B}u^{\mu}\chi_{+}\gamma_{5}\gamma_{\mu}B+\mathrm{H.c.}$&&&&&30&$\frac{1}{3}F_{28}^{(3)}$&I&$-\frac{1}{6}\tilde{C}_{44}^{(3)}$\\$i\bar{B}u^{\mu}\chi_{-}D_{\mu}B+\mathrm{H.c.}$&21&$-f_{21}^{(3)}$&I&$-\frac{1}{2}\tilde{c}_{9}^{(3)}$&31&$-F_{29}^{(3)}$&I&$-\frac{1}{2}\tilde{C}_{9}^{(3)}$\\$i\bar{B}\left\langle \nabla^{\mu}\chi_{-}\right\rangle \gamma_{5}\gamma_{\mu}B$&22&$\frac{1}{3}f_{22}^{(3)}$&I&$-\frac{1}{6}\tilde{c}_{41}^{(3)}$&32&$\frac{1}{3}F_{30}^{(3)}$&I&$-\frac{1}{6}\tilde{C}_{49}^{(3)}$\\$i\bar{B}\nabla^{\mu}\chi_{-}\gamma_{5}\gamma_{\mu}B$&23&$\frac{1}{3}f_{23}^{(3)}$&I&$-\frac{1}{6}\tilde{c}_{40}^{(3)}$&33&$\frac{1}{3}F_{31}^{(3)}$&I&$-\frac{1}{6}\tilde{C}_{48}^{(3)}$\\$\bar{T}^{\mu}\left\langle u_{\mu}u^{\nu}\right\rangle u^{\lambda}\gamma_{5}\gamma_{\nu}T_{\lambda}+\mathrm{H.c.}$&24&$f_{4}^{(3)}$&$-3\tilde{e}_{4}^{(3)}$&$*$&34&$*$&$*$&$*$\\$\bar{T}^{\mu}\left\langle u_{\mu}u^{\nu}\right\rangle u^{\lambda}\gamma_{5}\gamma_{\lambda}T_{\nu}$&25&$-2f_{4}^{(3)}$&$6\tilde{e}_{4}^{(3)}$&$*$&35&$*$&$*$&$*$\\$\bar{T}^{\mu}\left\langle u^{\nu}u_{\nu}\right\rangle u^{\lambda}\gamma_{5}\gamma_{\lambda}T_{\mu}$&26&$*$&$*$&$*$&36&$*$&$*$&$*$\\$\bar{T}^{\mu}\left\langle u^{\nu}u^{\lambda}\right\rangle u_{\nu}\gamma_{5}\gamma_{\lambda}T_{\mu}$&27&$*$&$*$&$*$&37&$*$&$*$&$*$\\$\bar{T}^{\mu}\left\langle u^{\nu}u^{\lambda}\right\rangle u^{\rho}\gamma_{5}\gamma_{\nu}D_{\lambda\rho}T_{\mu}$&28&$*$&$*$&$*$&38&$*$&$*$&$*$\\$\bar{T}^{\mu}\left\langle u_{\mu}u^{\nu}u^{\lambda}\right\rangle \gamma_{5}\gamma_{\nu}T_{\lambda}$&29&$-6f_{5}^{(3)}$&$-6\tilde{e}_{5}^{(3)}$&$6\tilde{c}_{1}^{(3)}$&39&$*$&$*$&$*$\\$\bar{T}^{\mu}\left\langle u_{\mu}u^{\nu}u^{\lambda}\right\rangle \gamma_{5}\gamma_{\lambda}T_{\nu}$&&&&&40&$*$&$*$&$*$\\$\bar{T}^{\mu}\left\langle u^{\nu}u_{\nu}u^{\lambda}\right\rangle \gamma_{5}\gamma_{\lambda}T_{\mu}$&&&&&41&$*$&$*$&$*$\\$\bar{T}^{\mu}\left\langle u^{\nu}u^{\lambda}u^{\rho}\right\rangle \gamma_{5}\gamma_{\nu}D_{\lambda\rho}T_{\mu}$&&&&&42&$*$&$*$&$*$\\$\bar{T}^{\mu}u_{\mu}u^{\nu}u^{\lambda}\gamma_{5}\gamma_{\nu}T_{\lambda}$&&&&&43&$-2F_{10}^{(3)}$&$-2\tilde{E}_{10}^{(3)}$&$\tilde{C}_{1}^{(3)}$\\$\bar{T}^{\mu}u_{\mu}u^{\nu}u^{\lambda}\gamma_{5}\gamma_{\lambda}T_{\nu}+\mathrm{H.c.}$&&&&&44&$-F_{8}^{(3)}$&$3\tilde{E}_{8}^{(3)}$&$*$\\$\bar{T}^{\mu}u^{\nu}u_{\mu}u^{\lambda}\gamma_{5}\gamma_{\lambda}T_{\nu}+\mathrm{H.c.}$&&&&&45&$*$&$*$&$*$\\$\bar{T}^{\mu}u^{\nu}u_{\nu}u^{\lambda}\gamma_{5}\gamma_{\lambda}T_{\mu}+\mathrm{H.c.}$&&&&&46&$*$&$*$&$*$\\$\bar{T}^{\mu}\left\langle u_{\mu}{f_{-}}^{\nu\lambda}\right\rangle D_{\nu}T_{\lambda}+\mathrm{H.c.}$&30&$*$&$*$&$*$&47&$*$&$*$&$*$\\$\bar{T}^{\mu}\left\langle u^{\nu}{f_{-\mu}}^{\lambda}\right\rangle D_{\nu}T_{\lambda}$&31&$*$&$*$&$*$&48&$*$&$*$&$*$\\$\bar{T}^{\mu}u_{\mu}{f_{-}}^{\nu\lambda}D_{\nu}T_{\lambda}+\mathrm{H.c.}$&32&$0$&$0$&$0$&49&$*$&$*$&$*$\\$\bar{T}^{\mu}u^{\nu}{f_{-\mu}}^{\lambda}D_{\nu}T_{\lambda}+\mathrm{H.c.}$&&&&&50&$*$&$*$&$*$\\$\bar{T}^{\mu}u^{\nu}{f_{-\mu}}^{\lambda}D_{\lambda}T_{\nu}+\mathrm{H.c.}$&&&&&51&$*$&$*$&$*$\\$\bar{T}^{\mu}u^{\nu}{f_{-\nu}}^{\lambda}D_{\lambda}T_{\mu}+\mathrm{H.c.}$&33&$-f_{15}^{(3)}$&$-\tilde{e}_{9}^{(3)}$&$\frac{1}{2}\tilde{c}_{2}^{(3)}$&52&$-F_{21}^{(3)}$&$-\tilde{E}_{14}^{(3)}$&$\frac{1}{2}\tilde{C}_{3}^{(3)}$\\$\bar{T}^{\mu}u_{\mu}h^{\nu\lambda}D_{\nu}T_{\lambda}+\mathrm{H.c.}$&34&$0$&$0$&$0$&53&$0$&$0$&$0$\\$\bar{T}^{\mu}u^{\nu}{h_{\nu}}^{\lambda}D_{\lambda}T_{\mu}+\mathrm{H.c.}$&35&$-f_{9}^{(3)}$&$-\tilde{e}_{10}^{(3)}$&$\frac{1}{2}\tilde{c}_{3}^{(3)}$&54&$-F_{14}^{(3)}$&$-\tilde{E}_{17}^{(3)}$&$\frac{1}{2}\tilde{C}_{4}^{(3)}$\\$\bar{T}^{\mu}u^{\nu}h^{\lambda\rho}D_{\nu\lambda\rho}T_{\mu}+\mathrm{H.c.}$&36&$f_{10}^{(3)}$&$-\tilde{e}_{11}^{(3)}$&$\frac{1}{2}\tilde{c}_{4}^{(3)}$&55&$F_{15}^{(3)}$&$-\tilde{E}_{18}^{(3)}$&$\frac{1}{2}\tilde{C}_{5}^{(3)}$\\$i\bar{T}^{\mu}\left\langle {f_{+\mu}}^{\nu}\right\rangle u^{\lambda}\gamma_{5}\gamma_{\nu}T_{\lambda}+\mathrm{H.c.}$&37&$*$&$*$&$*$&56&$0$&$0$&$0$\\$i\bar{T}^{\mu}\left\langle {f_{+\mu}}^{\nu}\right\rangle u^{\lambda}\gamma_{5}\gamma_{\lambda}T_{\nu}$&38&$-2f_{11}^{(3)}$&$2\tilde{e}_{15}^{(3)}$&$-\tilde{c}_{6}^{(3)}$&57&$0$&$0$&$0$\\$i\bar{T}^{\mu}\left\langle {f_{+\mu}}^{\nu}u^{\lambda}\right\rangle \gamma_{5}\gamma_{\nu}T_{\lambda}+\mathrm{H.c.}$&39&$*$&$*$&$*$&58&$2F_{17}^{(3)}$&$-2\tilde{E}_{24}^{(3)}$&$\tilde{C}_{7}^{(3)}$\\$i\bar{T}^{\mu}\left\langle {f_{+\mu}}^{\nu}u^{\lambda}\right\rangle \gamma_{5}\gamma_{\lambda}T_{\nu}$&40&$-2f_{12}^{(3)}$&$2\tilde{e}_{16}^{(3)}$&$*$&59&$-2F_{17}^{(3)}$&$2\tilde{E}_{24}^{(3)}$&$-\tilde{C}_{7}^{(3)}$\\$i\bar{T}^{\mu}{f_{+\mu}}^{\nu}u^{\lambda}\gamma_{5}\gamma_{\nu}T_{\lambda}+\mathrm{H.c.}$&41&$2f_{8}^{(3)}$&$-6\tilde{e}_{14}^{(3)}$&$-\tilde{c}_{41}^{(3)}$&60&$*$&$*$&$*$\\$i\bar{T}^{\mu}{f_{+\mu}}^{\nu}u^{\lambda}\gamma_{5}\gamma_{\lambda}T_{\nu}+\mathrm{H.c.}$&&&&&61&$-2F_{18}^{(3)}$&$2\tilde{E}_{25}^{(3)}$&$-\tilde{C}_{6}^{(3)}$\\$i\bar{T}^{\mu}{f_{+}}^{\nu\lambda}u_{\mu}\gamma_{5}\gamma_{\nu}T_{\lambda}+\mathrm{H.c.}$&&&&&62&$*$&$*$&$*$\\$i\bar{T}^{\mu}{f_{+}}^{\nu\lambda}u_{\nu}\gamma_{5}\gamma_{\lambda}T_{\mu}+\mathrm{H.c.}$&42&$*$&$*$&$*$&63&$*$&$*$&$*$\\$i\bar{T}^{\mu}\left\langle \nabla^{\nu}{f_{+\nu}}^{\lambda}\right\rangle D_{\lambda}T_{\mu}$&43&$f_{16}^{(3)}$&$-\tilde{e}_{17}^{(3)}$&$\frac{1}{2}\tilde{c}_{8}^{(3)}$&64&$0$&$0$&$0$\\$i\bar{T}^{\mu}\nabla^{\nu}{f_{+\nu}}^{\lambda}D_{\lambda}T_{\mu}$&44&$f_{17}^{(3)}$&$-\tilde{e}_{18}^{(3)}$&$\frac{1}{2}\tilde{c}_{7}^{(3)}$&65&$F_{24}^{(3)}$&$-\tilde{E}_{27}^{(3)}$&$\frac{1}{2}\tilde{C}_{8}^{(3)}$\\$\bar{T}^{\mu}\left\langle u^{\nu}\chi_{+}\right\rangle \gamma_{5}\gamma_{\nu}T_{\mu}$&45&$f_{19}^{(3)}$&$3\tilde{e}_{19}^{(3)}$&$-\frac{1}{2}\tilde{c}_{38}^{(3)}$&66&$F_{26}^{(3)}$&$3\tilde{E}_{28}^{(3)}$&$-\frac{1}{2}\tilde{C}_{45}^{(3)}$\\$\bar{T}^{\mu}\left\langle \chi_{+}\right\rangle u^{\nu}\gamma_{5}\gamma_{\nu}T_{\mu}$&46&$f_{20}^{(3)}$&$3\tilde{e}_{20}^{(3)}$&$*$&67&$F_{27}^{(3)}$&$3\tilde{E}_{29}^{(3)}$&$-\frac{1}{2}\tilde{C}_{46}^{(3)}$\\$\bar{T}^{\mu}u^{\nu}\chi_{+}\gamma_{5}\gamma_{\nu}T_{\mu}+\mathrm{H.c.}$&&&&&68&$F_{28}^{(3)}$&$3\tilde{E}_{30}^{(3)}$&$-\frac{1}{2}\tilde{C}_{44}^{(3)}$\\$i\bar{T}^{\mu}u^{\nu}\chi_{-}D_{\nu}T_{\mu}+\mathrm{H.c.}$&47&$f_{21}^{(3)}$&$-\tilde{e}_{21}^{(3)}$&$\frac{1}{2}\tilde{c}_{9}^{(3)}$&69&$F_{29}^{(3)}$&$-\tilde{E}_{31}^{(3)}$&$\frac{1}{2}\tilde{C}_{9}^{(3)}$\\$i\bar{T}^{\mu}\left\langle \nabla^{\nu}\chi_{-}\right\rangle \gamma_{5}\gamma_{\nu}T_{\mu}$&48&$f_{22}^{(3)}$&$3\tilde{e}_{22}^{(3)}$&$-\frac{1}{2}\tilde{c}_{41}^{(3)}$&70&$F_{30}^{(3)}$&$3\tilde{E}_{32}^{(3)}$&$-\frac{1}{2}\tilde{C}_{49}^{(3)}$\\$i\bar{T}^{\mu}\nabla^{\nu}\chi_{-}\gamma_{5}\gamma_{\nu}T_{\mu}$&49&$f_{23}^{(3)}$&$3\tilde{e}_{23}^{(3)}$&$-\frac{1}{2}\tilde{c}_{40}^{(3)}$&71&$F_{31}^{(3)}$&$3\tilde{E}_{33}^{(3)}$&$-\frac{1}{2}\tilde{C}_{48}^{(3)}$\\$\bar{B}\left\langle u^{\mu}u_{\mu}\right\rangle u^{\nu}T_{\nu}+\mathrm{H.c.}$&50&$\frac{2}{\sqrt{3}}f_{1}^{(3)}$&$2\sqrt{3}\tilde{e}_{1}^{(3)}$&$*$&72&$*$&$*$&$*$\\$\bar{B}\left\langle u^{\mu}u^{\nu}\right\rangle u_{\mu}T_{\nu}+\mathrm{H.c.}$&51&$\frac{2}{\sqrt{3}}f_{2}^{(3)}$&$2\sqrt{3}\tilde{e}_{2}^{(3)}$&$-\frac{1}{2\sqrt{3}}\tilde{c}_{10}^{(3)}$&73&$*$&$*$&$*$\\$\bar{B}\left\langle u^{\mu}u^{\nu}\right\rangle u^{\lambda}D_{\mu\nu}T_{\lambda}+\mathrm{H.c.}$&52&$*$&$*$&$*$&74&$*$&$*$&$*$\\$\bar{B}\left\langle u^{\mu}u^{\nu}\right\rangle u^{\lambda}D_{\mu\lambda}T_{\nu}+\mathrm{H.c.}$&53&$*$&$*$&$*$&75&$*$&$*$&$*$\\$i\bar{B}\left\langle u^{\mu}u^{\nu}\right\rangle u^{\lambda}\sigma_{\mu\lambda}T_{\nu}+\mathrm{H.c.}$&54&$0$&$0$&$0$&76&$0$&$0$&$0$\\$\bar{B}\left\langle u^{\mu}u_{\mu}u^{\nu}\right\rangle T_{\nu}+\mathrm{H.c.}$&&&&&77&$*$&$*$&$*$\\$\bar{B}\left\langle u^{\mu}u^{\nu}u^{\lambda}\right\rangle D_{\mu\nu}T_{\lambda}+\mathrm{H.c.}$&&&&&78&$*$&$*$&$*$\\$\bar{B}u^{\mu}u_{\mu}u^{\nu}T_{\nu}+\mathrm{H.c.}$&&&&&79&$0$&$0$&$0$\\$\bar{B}u^{\mu}u^{\nu}u_{\mu}T_{\nu}+\mathrm{H.c.}$&&&&&80&$-\frac{2}{\sqrt{3}}F_{7}^{(3)}$&$-2\sqrt{3}\tilde{E}_{7}^{(3)}$&$*$\\$\bar{B}u^{\mu}u^{\nu}u^{\lambda}D_{\mu\nu}T_{\lambda}+\mathrm{H.c.}$&&&&&81&$0$&$0$&$0$\\$\bar{B}u^{\mu}u^{\nu}u^{\lambda}D_{\mu\lambda}T_{\nu}+\mathrm{H.c.}$&&&&&82&$*$&$*$&$*$\\$i\bar{B}u^{\mu}u^{\nu}u^{\lambda}\sigma_{\mu\nu}T_{\lambda}+\mathrm{H.c.}$&&&&&83&$0$&$0$&$0$\\$i\bar{B}u^{\mu}u^{\nu}u^{\lambda}\sigma_{\mu\lambda}T_{\nu}+\mathrm{H.c.}$&&&&&84&$0$&$0$&$0$\\$\epsilon^{\mu\nu\lambda\rho}\bar{B}\left\langle u_{\mu}f_{-\nu\lambda}\right\rangle T_{\rho}+\mathrm{H.c.}$&55&$*$&$*$&$*$&85&$*$&$*$&$*$\\$\epsilon^{\mu\nu\lambda\rho}\bar{B}u_{\mu}f_{-\nu\lambda}T_{\rho}+\mathrm{H.c.}$&56&$0$&$0$&$0$&86&$*$&$*$&$*$\\$\epsilon^{\mu\nu\lambda\rho}\bar{B}f_{-\mu\nu}u_{\lambda}T_{\rho}+\mathrm{H.c.}$&&&&&87&$-\frac{2}{\sqrt{3}}F_{23}^{(3)}$&$-2\sqrt{3}\tilde{E}_{16}^{(3)}$&$-\frac{1}{2\sqrt{3}}\tilde{C}_{22}^{(3)}$\\$\bar{B}\left\langle u^{\mu}{f_{-}}^{\nu\lambda}\right\rangle \gamma_{5}\gamma_{\mu}D_{\nu}T_{\lambda}+\mathrm{H.c.}$&57&$\frac{4}{\sqrt{3}}f_{6}^{(3)}$&$4\sqrt{3}\tilde{e}_{8}^{(3)}$&$\frac{2}{\sqrt{3}}\tilde{c}_{19}^{(3)}$&88&$\frac{4}{\sqrt{3}}F_{11}^{(3)}$&$4\sqrt{3}\tilde{E}_{13}^{(3)}$&$\frac{2}{\sqrt{3}}\tilde{C}_{32}^{(3)}$\\$\bar{B}\left\langle u^{\mu}{f_{-}}^{\nu\lambda}\right\rangle \gamma_{5}\gamma_{\nu}D_{\mu}T_{\lambda}+\mathrm{H.c.}$&58&$*$&$*$&$*$&89&$*$&$*$&$*$\\$\bar{B}\left\langle u^{\mu}h^{\nu\lambda}\right\rangle \gamma_{5}\gamma_{\mu}D_{\nu}T_{\lambda}+\mathrm{H.c.}$&59&$-\frac{4}{\sqrt{3}}f_{6}^{(3)}$&$-4\sqrt{3}\tilde{e}_{8}^{(3)}$&$-\frac{2}{\sqrt{3}}\tilde{c}_{19}^{(3)}$&90&$-\frac{4}{\sqrt{3}}F_{11}^{(3)}$&$-4\sqrt{3}\tilde{E}_{13}^{(3)}$&$-\frac{2}{\sqrt{3}}\tilde{C}_{32}^{(3)}$\\$\bar{B}u^{\mu}{f_{-}}^{\nu\lambda}\gamma_{5}\gamma_{\mu}D_{\nu}T_{\lambda}+\mathrm{H.c.}$&60&$0$&$0$&$0$&91&$*$&$*$&$*$\\$\bar{B}u^{\mu}{f_{-}}^{\nu\lambda}\gamma_{5}\gamma_{\nu}D_{\mu}T_{\lambda}+\mathrm{H.c.}$&61&$0$&$0$&$0$&92&$*$&$*$&$*$\\$\bar{B}u^{\mu}h^{\nu\lambda}\gamma_{5}\gamma_{\mu}D_{\nu}T_{\lambda}+\mathrm{H.c.}$&62&$0$&$0$&$0$&93&$-\frac{4}{\sqrt{3}}F_{16}^{(3)}$&$-4\sqrt{3}\tilde{E}_{19}^{(3)}$&$-\frac{2}{\sqrt{3}}\tilde{C}_{25}^{(3)}$\\$\bar{B}u^{\mu}h^{\nu\lambda}\gamma_{5}\gamma_{\nu}D_{\mu}T_{\lambda}+\mathrm{H.c.}$&&&&&94&$*$&$*$&$*$\\$\bar{B}u^{\mu}h^{\nu\lambda}\gamma_{5}\gamma_{\nu}D_{\lambda}T_{\mu}+\mathrm{H.c.}$&&&&&95&$*$&$*$&$*$\\$\bar{B}{f_{-}}^{\mu\nu}u^{\lambda}\gamma_{5}\gamma_{\mu}D_{\nu}T_{\lambda}+\mathrm{H.c.}$&&&&&96&$*$&$*$&$*$\\$\bar{B}\nabla^{\mu}{f_{-\mu}}^{\nu}T_{\nu}+\mathrm{H.c.}$&63&$\frac{2}{\sqrt{3}}f_{18}^{(3)}$&$2\sqrt{3}\tilde{e}_{12}^{(3)}$&$\frac{1}{\sqrt{3}}\tilde{c}_{24}^{(3)}$&97&$\frac{2}{\sqrt{3}}F_{25}^{(3)}$&$2\sqrt{3}\tilde{E}_{20}^{(3)}$&$\frac{1}{\sqrt{3}}\tilde{C}_{33}^{(3)}$\\$i\bar{B}\left\langle {f_{+}}^{\mu\nu}\right\rangle u_{\mu}T_{\nu}+\mathrm{H.c.}$&64&$-\frac{2}{\sqrt{3}}f_{7}^{(3)}$&$-2\sqrt{3}\tilde{e}_{13}^{(3)}$&$-\frac{1}{\sqrt{3}}\tilde{c}_{30}^{(3)}$&98&$0$&$0$&$0$\\$i\bar{B}\left\langle {f_{+}}^{\mu\nu}\right\rangle u^{\lambda}D_{\mu\lambda}T_{\nu}+\mathrm{H.c.}$&65&$0$&$0$&$0$&99&$0$&$0$&$0$\\$\bar{B}\left\langle {f_{+}}^{\mu\nu}\right\rangle u^{\lambda}\sigma_{\mu\nu}T_{\lambda}+\mathrm{H.c.}$&66&$0$&$0$&$0$&100&$0$&$0$&$0$\\$i\bar{B}\left\langle {f_{+}}^{\mu\nu}u_{\mu}\right\rangle T_{\nu}+\mathrm{H.c.}$&67&$-\frac{2}{\sqrt{3}}f_{7}^{(3)}$&$-2\sqrt{3}\tilde{e}_{13}^{(3)}$&$-\frac{1}{\sqrt{3}}\tilde{c}_{30}^{(3)}$&101&$0$&$0$&$0$\\$i\bar{B}\left\langle {f_{+}}^{\mu\nu}u^{\lambda}\right\rangle D_{\mu\lambda}T_{\nu}+\mathrm{H.c.}$&68&$0$&$0$&$0$&102&$0$&$0$&$0$\\$\bar{B}\left\langle {f_{+}}^{\mu\nu}u^{\lambda}\right\rangle \sigma_{\mu\nu}T_{\lambda}+\mathrm{H.c.}$&69&$0$&$0$&$0$&103&$0$&$0$&$0$\\$i\bar{B}{f_{+}}^{\mu\nu}u_{\mu}T_{\nu}+\mathrm{H.c.}$&70&$\frac{4}{\sqrt{3}}f_{7}^{(3)}$&$4\sqrt{3}\tilde{e}_{13}^{(3)}$&$\frac{2}{\sqrt{3}}\tilde{c}_{30}^{(3)}$&104&$\frac{2}{\sqrt{3}}F_{12}^{(3)}$&$2\sqrt{3}\tilde{E}_{21}^{(3)}$&$\frac{1}{\sqrt{3}}\tilde{C}_{39}^{(3)}$\\$i\bar{B}{f_{+}}^{\mu\nu}u^{\lambda}D_{\mu\lambda}T_{\nu}+\mathrm{H.c.}$&&&&&105&$0$&$0$&$0$\\$\bar{B}{f_{+}}^{\mu\nu}u^{\lambda}\sigma_{\mu\nu}T_{\lambda}+\mathrm{H.c.}$&71&$0$&$0$&$0$&106&$0$&$0$&$0$\\$i\bar{B}u^{\mu}{f_{+\mu}}^{\nu}T_{\nu}+\mathrm{H.c.}$&&&&&107&$-\frac{2}{\sqrt{3}}F_{12}^{(3)}$&$-2\sqrt{3}\tilde{E}_{21}^{(3)}$&$-\frac{1}{\sqrt{3}}\tilde{C}_{39}^{(3)}$\\$\bar{B}u^{\mu}{f_{+}}^{\nu\lambda}\sigma_{\mu\nu}T_{\lambda}+\mathrm{H.c.}$&&&&&108&$0$&$0$&$0$\\$\bar{B}\left\langle u^{\mu}\chi_{+}\right\rangle T_{\mu}+\mathrm{H.c.}$&72&$\frac{2}{\sqrt{3}}f_{19}^{(3)}$&$2\sqrt{3}\tilde{e}_{19}^{(3)}$&$-\frac{1}{\sqrt{3}}\tilde{c}_{38}^{(3)}$&109&$\frac{2}{\sqrt{3}}F_{26}^{(3)}$&$2\sqrt{3}\tilde{E}_{28}^{(3)}$&$-\frac{1}{\sqrt{3}}\tilde{C}_{45}^{(3)}$\\$\bar{B}\left\langle \chi_{+}\right\rangle u^{\mu}T_{\mu}+\mathrm{H.c.}$&73&$\frac{2}{\sqrt{3}}f_{20}^{(3)}$&$2\sqrt{3}\tilde{e}_{20}^{(3)}$&$*$&110&$\frac{2}{\sqrt{3}}F_{27}^{(3)}$&$2\sqrt{3}\tilde{E}_{29}^{(3)}$&$-\frac{1}{\sqrt{3}}\tilde{C}_{46}^{(3)}$\\$\bar{B}u^{\mu}\chi_{+}T_{\mu}+\mathrm{H.c.}$&74&$0$&$0$&$0$&111&$\frac{2}{\sqrt{3}}F_{28}^{(3)}$&$2\sqrt{3}\tilde{E}_{30}^{(3)}$&$-\frac{1}{\sqrt{3}}\tilde{C}_{44}^{(3)}$\\$\bar{B}\chi_{+}u^{\mu}T_{\mu}+\mathrm{H.c.}$&&&&&112&$\frac{2}{\sqrt{3}}F_{28}^{(3)}$&$2\sqrt{3}\tilde{E}_{30}^{(3)}$&$-\frac{1}{\sqrt{3}}\tilde{C}_{44}^{(3)}$\\
\end{longtable}

\begin{longtable}{lcclcc}
\caption{Independent Lagrangian terms in the heavy diquark limit at the order $\mathcal{O}(p^{3})$. The second and fifth (third and sixth) columns label the number for each term in the two (three)-flavor case. The terms without a number are not independent.}\label{tab:table6}\\
	%\begin{ruledtabular}
	\hline\hline $P_n/p_n$  & $SU(2)$ & $SU(3)$&$P_n/p_n$  & $SU(2)$ & $SU(3)$\\
	\hline%\addlinespace[2pt]
	\endfirsthead
	\hline\hline $P_n/p_n$  & $SU(2)$ & $SU(3)$&$P_n/p_n$  & $SU(2)$ & $SU(3)$\\
	\hline%\addlinespace[2pt]
	\endhead
	\hline\hline
	\endfoot
	\hline\hline
	\endlastfoot
	$\psib^{\mu}\la u^{\nu}u_{\nu}\ra u^{\lambda}\gamf\gamma_{\lambda}\psi_{\mu}$ & 1 & 1 & $\epsilon^{\mu\nu\lambda\rho}\psib^{\sigma}\la f_{+\mu\nu}u_{\lambda}\ra v_{\rho}\psi_{\sigma}$ & 12 & 17  \\
	$\psib^{\mu}\la u^{\nu}u^{\lambda}\ra u_{\nu}\gamf\gamma_{\lambda}\psi_{\mu}$ & 2 & 2 & $\epsilon^{\mu\nu\lambda\rho}\psib^{\sigma}f_{+\mu\nu}u_{\lambda}v_{\rho}\psi_{\sigma}+\mathrm{H.c.}$ &  & 18  \\
	$\psib^{\mu}\la u^{\nu}u^{\lambda}\ra u^{\rho}\gamf\gamma_{\nu}v_{\lambda}v_{\rho}\psi_{\mu}$ & 3 & 3 & $\psib^{\mu}\la u^{\nu}{f_{-}}^{\lambda\rho}\ra\sigma_{\nu\lambda}v_{\rho}\psi_{\mu}$ & 13 & 19  \\
	$\psib^{\mu}\la u^{\nu}u^{\lambda}\ra u^{\rho}\gamf\gamma_{\rho}v_{\nu}v_{\lambda}\psi_{\mu}$ & 4 & 4 & $\psib^{\mu}\la u^{\nu}{f_{-}}^{\lambda\rho}\ra\sigma_{\lambda\rho}v_{\nu}\psi_{\mu}$ & 14 & 20  \\
	$\psib^{\mu}\la u^{\nu}u_{\nu}u^{\lambda}\ra\gamf\gamma_{\lambda}\psi_{\mu}$ &  & 5 & $i\psib^{\mu}u^{\nu}{f_{-\nu}}^{\lambda}v_{\lambda}\psi_{\mu}+\mathrm{H.c.}$ & 15 & 21  \\
	$\psib^{\mu}\la u^{\nu}u^{\lambda}u^{\rho}\ra\gamf\gamma_{\nu}v_{\lambda}v_{\rho}\psi_{\mu}$ &  & 6 & $\psib^{\mu}\la\nabla^{\nu}{f_{+\nu}}^{\lambda}\ra v_{\lambda}\psi_{\mu}$ & 16 &  \\
	$\psib^{\mu}u^{\nu}u_{\nu}u^{\lambda}\gamf\gamma_{\lambda}\psi_{\mu}+\mathrm{H.c.}$ &  & 7 & $\psib^{\mu}u^{\nu}{f_{-}}^{\lambda\rho}\sigma_{\nu\lambda}v_{\rho}\psi_{\mu}+\mathrm{H.c.}$ &  & 22  \\
	$\psib^{\mu}u^{\nu}u^{\lambda}u^{\rho}\gamf\gamma_{\nu}v_{\lambda}v_{\rho}\psi_{\mu}+\mathrm{H.c.}$ &  & 8 & $\psib^{\mu}u^{\nu}{f_{-}}^{\lambda\rho}\sigma_{\lambda\rho}v_{\nu}\psi_{\mu}+\mathrm{H.c.}$ &  & 23  \\
	$i\epsilon^{\mu\nu\lambda\rho}\psib^{\sigma}\la u_{\mu}u_{\nu}u_{\lambda}\ra v_{\rho}\psi_{\sigma}$ & 5 & 9 & $\psib^{\mu}\nabla^{\nu}{f_{+\nu}}^{\lambda}v_{\lambda}\psi_{\mu}$ & 17 & 24  \\
	$i\epsilon^{\mu\nu\lambda\rho}\psib^{\sigma}u_{\mu}u_{\nu}u_{\lambda}v_{\rho}\psi_{\sigma}$ &  & 10 & $\psib^{\mu}\nabla^{\nu}{f_{-\nu}}^{\lambda}\gamf\gamma_{\lambda}\psi_{\mu}$ & 18 & 25  \\
	$\psib^{\mu}\la u^{\nu}h^{\lambda\rho}\ra\sigma_{\nu\lambda}v_{\rho}\psi_{\mu}$ & 6 & 11 & $\psib^{\mu}\la u^{\nu}\chip\ra\gamf\gamma_{\nu}\psi_{\mu}$ & 19 & 26  \\
	$i\psib^{\mu}{f_{+}}^{\nu\lambda}u_{\nu}\gamf\gamma_{\lambda}\psi_{\mu}+\mathrm{H.c.}$ & 7 & 12 & $\psib^{\mu}\la\chip\ra u^{\nu}\gamf\gamma_{\nu}\psi_{\mu}$ & 20 & 27  \\
	$i\psib^{\mu}{f_{+}}^{\nu\lambda}u^{\rho}\gamf\gamma_{\nu}v_{\lambda}v_{\rho}\psi_{\mu}+\mathrm{H.c.}$ & 8 & 13 & $\psib^{\mu}u^{\nu}\chip\gamf\gamma_{\nu}\psi_{\mu}+\mathrm{H.c.}$ &  & 28  \\
	$i\psib^{\mu}u^{\nu}{h_{\nu}}^{\lambda}v_{\lambda}\psi_{\mu}+\mathrm{H.c.}$ & 9 & 14 & $\psib^{\mu}u^{\nu}\chim v_{\nu}\psi_{\mu}+\mathrm{H.c.}$ & 21 & 29  \\
	$i\psib^{\mu}u^{\nu}h^{\lambda\rho}v_{\nu}v_{\lambda}v_{\rho}\psi_{\mu}+\mathrm{H.c.}$ & 10 & 15 & $i\psib^{\mu}\la\nabla^{\nu}\chi_{-}\ra\gamf\gamma_{\nu}\psi_{\mu}$ & 22 & 30  \\
	$\epsilon^{\mu\nu\lambda\rho}\psib^{\sigma}\la f_{+\mu\nu}\ra u_{\lambda}v_{\rho}\psi_{\sigma}$ & 11 & & $i\psib^{\mu}\nabla^{\nu}\chi_{-}\gamf\gamma_{\nu}\psi_{\mu}$ & 23 & 31  \\
	$\psib^{\mu}u^{\nu}h^{\lambda\rho}\sigma_{\nu\lambda}v_{\rho}\psi_{\mu}+\mathrm{H.c.}$ &  & 16  \\
	%\end{ruledtabular}
\end{longtable}

\begin{align}\label{eqA1}
\begin{autobreak}
~
\tilde{e}_ {26}^{(3)}=f_ {1}^{(3)}+f_ {4}^{(3)},\;
\tilde{e}_ {27}^{(3)}=f_ {2}^{(3)}-f_ {4}^{(3)},\;
\tilde{e}_ {28}^{(3)}=-f_{3}^{(3)}-f_ {4}^{(3)},\;
\tilde{e}_ {30}^{(3)}=f_ {13}^{(3)}+f_ {6}^{(3)},\;
\tilde{e}_ {31}^{(3)}=-2f_ {14}^{(3)}+2f_ {6}^{(3)},\;
\tilde{e}_ {37}^{(3)}=2f_ {11}^{(3)}-f_ {8}^{(3)},\;
\tilde{e}_ {39}^{(3)}=2f_ {12}^{(3)}-f_ {8}^{(3)},\;
\tilde{e}_ {42}^{(3)}=f_ {7}^{(3)}-f_ {8}^{(3)},\;
\tilde{e}_ {52}^{(3)}=-\frac{2}{\sqrt{3}}f_ {4}^{(3)}-\frac{2}{\sqrt{3}}f_ {8}^{(3)},\;
\tilde{e}_ {53}^{(3)}=-\frac{2}{\sqrt{3}}f_ {3}^{(3)}+\frac{2}{\sqrt{3}}f_ {8}^{(3)},\;
\tilde{e}_ {55}^{(3)}=-\frac{1}{\sqrt{3}}f_ {13}^{(3)}+\frac{1}{\sqrt{3}}f_ {6}^{(3)},\;
\tilde{e}_ {58}^{(3)}=\frac{2}{\sqrt{3}}f_ {13}^{(3)}-\frac{2}{\sqrt{3}}f_ {6}^{(3)}-\frac{4}{\sqrt{3}}f_ {14}^{(3)}.
\end{autobreak}
\end{align}

\begin{align}\label{eqA2}
\begin{autobreak}
~
\tilde{e}_{26}^{(3)}=3\tilde{e}_{1}^{(3)}-3\tilde{e}_{4}^{(3)},\;
\tilde{e}_{27}^{(3)}=3\tilde{e}_{2}^{(3)}+3\tilde{e}_{4}^{(3)},\;
\tilde{e}_{28}^{(3)}=3\tilde{e}_{3}^{(3)}+3\tilde{e}_{4}^{(3)},\;
\tilde{e}_{30}^{(3)}=3\tilde{e}_{6}^{(3)}+3\tilde{e}_{8}^{(3)},\;
\tilde{e}_{31}^{(3)}=-6\tilde{e}_{7}^{(3)}+6\tilde{e}_{8}^{(3)},\;
\tilde{e}_{37}^{(3)}=-2\tilde{e}_{15}^{(3)}+3\tilde{e}_{14}^{(3)},\;
\tilde{e}_{39}^{(3)}=-2\tilde{e}_{16}^{(3)}+3\tilde{e}_{14}^{(3)},\;
\tilde{e}_{42}^{(3)}=3\tilde{e}_{13}^{(3)}+3\tilde{e}_{14}^{(3)},\;
\tilde{e}_{52}^{(3)}=2\sqrt{3}\tilde{e}_{14}^{(3)}+2\sqrt{3}\tilde{e}_{4}^{(3)},\;
\tilde{e}_{53}^{(3)}=-2\sqrt{3}\tilde{e}_{14}^{(3)}+2\sqrt{3}\tilde{e}_{3}^{(3)},\;
\tilde{e}_{55}^{(3)}=-\sqrt{3}\tilde{e}_{6}^{(3)}+\sqrt{3}\tilde{e}_{8}^{(3)},\;
\tilde{e}_{58}^{(3)}=2\sqrt{3}\tilde{e}_{6}^{(3)}-2\sqrt{3}\tilde{e}_{8}^{(3)}-4\sqrt{3}\tilde{e}_{7}^{(3)}.
\end{autobreak}
\end{align}

\begin{align}\label{eqA3}
\begin{autobreak}
~
\tilde{e}_ {1}^{(3)}=-\frac{1}{12}\tilde{c}_ {13}^{(3)}+\frac{1}{24}\tilde{c}_ {10}^{(3)},\
\tilde{e}_ {3}^{(3)}=-\frac{1}{12}\tilde{c}_ {10}^{(3)}-\frac{1}{6}\tilde{c}_ {11}^{(3)}-\frac{1}{6}\tilde{c}_ {14}^{(3)},\;
\tilde{e}_ {4}^{(3)}=\frac{1}{3}\tilde{c}_ {11}^{(3)}-\frac{1}{3}\tilde{c}_ {14}^{(3)}+\frac{1}{6}\tilde{c}_ {10}^{(3)},\;
\tilde{e}_ {16}^{(3)}=\frac{1}{2}\tilde{c}_ {5}^{(3)}-\frac{1}{2}\tilde{c}_ {6}^{(3)},\;
\tilde{e}_ {20}^{(3)}=\frac{1}{6}\tilde{c}_ {37}^{(3)}-\frac{1}{6}\tilde{c}_ {38}^{(3)},\;
\tilde{e}_ {24}^{(3)}=-\frac{1}{2}\tilde{c}_ {10}^{(3)}-\tilde{c}_ {11}^{(3)}+\tilde{c}_ {14}^{(3)},\;
\tilde{e}_ {25}^{(3)}=2\tilde{c}_ {11}^{(3)}-2\tilde{c}_ {14}^{(3)}+\tilde{c}_ {10}^{(3)},\;
\tilde{e}_ {26}^{(3)}=-\frac{1}{4}\tilde{c}_ {13}^{(3)}-\frac{3}{8}\tilde{c}_ {10}^{(3)}-\tilde{c}_ {11}^{(3)}+\tilde{c}_ {14}^{(3)},\;
\tilde{e}_ {27}^{(3)}=\frac{1}{4}\tilde{c}_ {10}^{(3)}+\tilde{c}_ {11}^{(3)}-\tilde{c}_ {14}^{(3)},\;
\tilde{e}_ {28}^{(3)}=\frac{1}{2}\tilde{c}_ {11}^{(3)}+\frac{1}{4}\tilde{c}_ {10}^{(3)}-\frac{3}{2}\tilde{c}_ {14}^{(3)},\;
\tilde{e}_ {30}^{(3)}=-\frac{1}{2}\tilde{c}_ {15}^{(3)}+\frac{1}{2}\tilde{c}_ {19}^{(3)},\;
\tilde{e}_ {31}^{(3)}=-\frac{1}{2}\tilde{c}_ {16}^{(3)}+\tilde{c}_ {19}^{(3)},\;
\tilde{e}_ {37}^{(3)}=\frac{1}{2}\tilde{c}_ {41}^{(3)}+\tilde{c}_ {6}^{(3)},\;
\tilde{e}_ {39}^{(3)}=\frac{1}{2}\tilde{c}_ {41}^{(3)}-\tilde{c}_ {5}^{(3)}+\tilde{c}_ {6}^{(3)},\;
\tilde{e}_ {40}^{(3)}=\tilde{c}_ {5}^{(3)}-\tilde{c}_ {6}^{(3)},\;
\tilde{e}_ {42}^{(3)}=\frac{1}{2}\tilde{c}_ {30}^{(3)}+\frac{1}{2}\tilde{c}_ {41}^{(3)},\;
\tilde{e}_ {46}^{(3)}=\frac{1}{2}\tilde{c}_ {37}^{(3)}-\frac{1}{2}\tilde{c}_ {38}^{(3)},\;
\tilde{e}_ {50}^{(3)}=-\frac{1}{2\sqrt{3}}\tilde{c}_ {13}^{(3)}+\frac{1}{4\sqrt{3}}\tilde{c}_ {10}^{(3)},\;
\tilde{e}_ {52}^{(3)}=\frac{1}{\sqrt{3}}\tilde{c}_ {10}^{(3)}+\frac{1}{\sqrt{3}}\tilde{c}_ {41}^{(3)}+\frac{2}{\sqrt{3}}\tilde{c}_ {11}^{(3)}-\frac{2}{\sqrt{3}}\tilde{c}_ {14}^{(3)},\;
\tilde{e}_ {53}^{(3)}=-\frac{1}{2\sqrt{3}}\tilde{c}_ {10}^{(3)}-\frac{1}{\sqrt{3}}\tilde{c}_ {11}^{(3)}-\frac{1}{\sqrt{3}}\tilde{c}_ {14}^{(3)}-\frac{1}{\sqrt{3}}\tilde{c}_ {41}^{(3)},\;
\tilde{e}_ {55}^{(3)}=\frac{1}{2\sqrt{3}}\tilde{c}_ {15}^{(3)}+\frac{1}{2\sqrt{3}}\tilde{c}_ {19}^{(3)},\;
\tilde{e}_ {58}^{(3)}=-\frac{1}{\sqrt{3}}\tilde{c}_ {15}^{(3)}-\frac{1}{\sqrt{3}}\tilde{c}_ {16}^{(3)}-\frac{1}{\sqrt{3}}\tilde{c}_ {19}^{(3)},\;
\tilde{e}_ {73}^{(3)}=\frac{1}{\sqrt{3}}\tilde{c}_ {37}^{(3)}-\frac{1}{\sqrt{3}}\tilde{c}_ {38}^{(3)}.
\end{autobreak}
\end{align}

\begin{align}\label{eqA4}
\begin{autobreak}
~
\tilde{E}_ {34}^{(3)}=F_ {10}^{(3)}+F_ {4}^{(3)}+F_ {8}^{(3)},\;
\tilde{E}_ {35}^{(3)}=-2F_ {4}^{(3)}+F_ {10}^{(3)},\;
\tilde{E}_ {36}^{(3)}=F_ {1}^{(3)}+F_ {4}^{(3)},\;
\tilde{E}_ {37}^{(3)}=F_ {2}^{(3)}-F_ {4}^{(3)}-F_ {8}^{(3)},\;
\tilde{E}_ {38}^{(3)}=-F_{3}^{(3)}-F_ {4}^{(3)}-F_ {8}^{(3)},\;
\tilde{E}_ {39}^{(3)}=-3F_ {9}^{(3)}+F_ {10}^{(3)}+\frac{2}{3}F_ {8}^{(3)},\;
\tilde{E}_ {40}^{(3)}=3F_ {9}^{(3)}+F_ {10}^{(3)}+\frac{2}{3}F_ {8}^{(3)},\;
\tilde{E}_ {41}^{(3)}=F_ {5}^{(3)}-\frac{2}{3}F_ {8}^{(3)},\;
\tilde{E}_ {42}^{(3)}=-F_{6}^{(3)}-\frac{2}{3}F_ {8}^{(3)},\;
\tilde{E}_ {45}^{(3)}=-2F_ {10}^{(3)}-F_ {8}^{(3)},\;
\tilde{E}_ {46}^{(3)}=F_ {7}^{(3)}+F_ {8}^{(3)},\;
\tilde{E}_ {47}^{(3)}=F_ {11}^{(3)}+F_ {19}^{(3)},\;
\tilde{E}_ {48}^{(3)}=2F_ {11}^{(3)}-2F_ {20}^{(3)},\;
\tilde{E}_ {49}^{(3)}=F_ {16}^{(3)}+F_ {22}^{(3)},\;
\tilde{E}_ {50}^{(3)}=2F_ {16}^{(3)}-2F_ {23}^{(3)},\;
\tilde{E}_ {51}^{(3)}=F_ {16}^{(3)}+F_ {22}^{(3)},\;
\tilde{E}_ {60}^{(3)}=2F_ {18}^{(3)}+F_ {13}^{(3)},\;
\tilde{E}_ {62}^{(3)}=2F_ {18}^{(3)}-F_ {13}^{(3)},\;
\tilde{E}_ {63}^{(3)}=F_ {12}^{(3)}-F_ {13}^{(3)},\;
\tilde{E}_ {72}^{(3)}=\frac{1}{\sqrt{3}}F_ {7}^{(3)}+\frac{2}{\sqrt{3}}F_ {1}^{(3)},\;
\tilde{E}_ {73}^{(3)}=\frac{2}{\sqrt{3}}F_ {2}^{(3)}+\frac{2}{\sqrt{3}}F_ {7}^{(3)},\;
\tilde{E}_ {74}^{(3)}=-\frac{1}{2\sqrt{3}}F_ {13}^{(3)}-\frac{1}{\sqrt{3}}F_ {8}^{(3)}-\frac{2}{\sqrt{3}}F_ {4}^{(3)},\;
\tilde{E}_ {75}^{(3)}=-\frac{1}{\sqrt{3}}F_ {13}^{(3)}-\frac{2}{\sqrt{3}}F_ {3}^{(3)}-\frac{2}{\sqrt{3}}F_ {8}^{(3)},\;
\tilde{E}_ {77}^{(3)}=\frac{2}{\sqrt{3}}F_ {5}^{(3)}+\frac{2}{\sqrt{3}}F_ {7}^{(3)},\;
\tilde{E}_ {78}^{(3)}=-\frac{1}{\sqrt{3}}F_ {13}^{(3)}-\frac{2}{\sqrt{3}}F_ {6}^{(3)}-\frac{2}{\sqrt{3}}F_ {8}^{(3)},\;
\tilde{E}_ {82}^{(3)}=\frac{2}{\sqrt{3}}F_ {8}^{(3)}+\sqrt{3}F_ {13}^{(3)},\;
\tilde{E}_ {85}^{(3)}=\frac{1}{\sqrt{3}}F_ {11}^{(3)}-\frac{1}{\sqrt{3}}F_ {19}^{(3)},\;
\tilde{E}_ {86}^{(3)}=-\frac{1}{2\sqrt{3}}F_ {16}^{(3)}+\frac{1}{\sqrt{3}}F_ {23}^{(3)}-\frac{\sqrt{3}}{2}F_ {22}^{(3)},\;
\tilde{E}_ {89}^{(3)}=-\frac{2}{\sqrt{3}}F_ {11}^{(3)}+\frac{2}{\sqrt{3}}F_ {19}^{(3)}-\frac{4}{\sqrt{3}}F_ {20}^{(3)},\;
\tilde{E}_ {91}^{(3)}=\frac{1}{\sqrt{3}}F_ {16}^{(3)}-\frac{1}{\sqrt{3}}F_ {22}^{(3)}+\frac{2}{\sqrt{3}}F_ {23}^{(3)},\;
\tilde{E}_ {92}^{(3)}=\frac{2}{\sqrt{3}}F_ {16}^{(3)}+\frac{2}{\sqrt{3}}F_ {22}^{(3)}-\frac{4}{\sqrt{3}}F_ {23}^{(3)},\;
\tilde{E}_ {94}^{(3)}=\frac{1}{\sqrt{3}}F_ {16}^{(3)}-\frac{1}{\sqrt{3}}F_ {22}^{(3)}+\frac{2}{\sqrt{3}}F_ {23}^{(3)},\;
\tilde{E}_ {95}^{(3)}=\frac{1}{\sqrt{3}}F_ {22}^{(3)}-\frac{2}{\sqrt{3}}F_ {23}^{(3)}+\sqrt{3}F_ {16}^{(3)},\;
\tilde{E}_ {96}^{(3)}=-\frac{2}{\sqrt{3}}F_ {16}^{(3)}+\frac{2}{\sqrt{3}}F_ {22}^{(3)}-\frac{4}{\sqrt{3}}F_ {23}^{(3)}.
\end{autobreak}
\end{align}

\begin{align}\label{eqA5}
\begin{autobreak}
~
\tilde{E}_ {34}^{(3)}=-3\tilde{E}_ {4}^{(3)}-3\tilde{E}_ {8}^{(3)}+\tilde{E}_ {10}^{(3)},\;
\tilde{E}_ {35}^{(3)}=6\tilde{E}_ {4}^{(3)}+\tilde{E}_ {10}^{(3)},\;
\tilde{E}_ {36}^{(3)}=3\tilde{E}_ {1}^{(3)}-3\tilde{E}_ {4}^{(3)},\;
\tilde{E}_ {37}^{(3)}=3\tilde{E}_ {2}^{(3)}+3\tilde{E}_ {4}^{(3)}+3\tilde{E}_ {8}^{(3)},\;
\tilde{E}_ {38}^{(3)}=3\tilde{E}_ {3}^{(3)}+3\tilde{E}_ {4}^{(3)}+3\tilde{E}_ {8}^{(3)},\;
\tilde{E}_ {39}^{(3)}=-2\tilde{E}_ {8}^{(3)}-3\tilde{E}_ {9}^{(3)}+\tilde{E}_ {10}^{(3)},\;
\tilde{E}_ {40}^{(3)}=-2\tilde{E}_ {8}^{(3)}+3\tilde{E}_ {9}^{(3)}+\tilde{E}_ {10}^{(3)},\;
\tilde{E}_ {41}^{(3)}=2\tilde{E}_ {8}^{(3)}+3\tilde{E}_ {5}^{(3)},\;
\tilde{E}_ {42}^{(3)}=2\tilde{E}_ {8}^{(3)}+3\tilde{E}_ {6}^{(3)},\;
\tilde{E}_ {45}^{(3)}=-2\tilde{E}_ {10}^{(3)}+3\tilde{E}_ {8}^{(3)},\;
\tilde{E}_ {46}^{(3)}=3\tilde{E}_ {7}^{(3)}-3\tilde{E}_ {8}^{(3)},\;
\tilde{E}_ {47}^{(3)}=3\tilde{E}_ {11}^{(3)}+3\tilde{E}_ {13}^{(3)},\;
\tilde{E}_ {48}^{(3)}=-6\tilde{E}_ {12}^{(3)}+6\tilde{E}_ {13}^{(3)},\;
\tilde{E}_ {49}^{(3)}=3\tilde{E}_ {15}^{(3)}+3\tilde{E}_ {19}^{(3)},\;
\tilde{E}_ {50}^{(3)}=-6\tilde{E}_ {16}^{(3)}+6\tilde{E}_ {19}^{(3)},\;
\tilde{E}_ {51}^{(3)}=3\tilde{E}_ {15}^{(3)}+3\tilde{E}_ {19}^{(3)},\;
\tilde{E}_ {60}^{(3)}=-2\tilde{E}_ {25}^{(3)}-3\tilde{E}_ {22}^{(3)},\;
\tilde{E}_ {62}^{(3)}=-2\tilde{E}_ {25}^{(3)}+3\tilde{E}_ {22}^{(3)},\;
\tilde{E}_ {63}^{(3)}=3\tilde{E}_ {21}^{(3)}+3\tilde{E}_ {22}^{(3)},\;
\tilde{E}_ {72}^{(3)}=2\sqrt{3}\tilde{E}_ {1}^{(3)}+\sqrt{3}\tilde{E}_ {7}^{(3)},\;
\tilde{E}_ {73}^{(3)}=2\sqrt{3}\tilde{E}_ {2}^{(3)}+2\sqrt{3}\tilde{E}_ {7}^{(3)},\;
\tilde{E}_ {74}^{(3)}=2\sqrt{3}\tilde{E}_ {4}^{(3)}+\frac{\sqrt{3}}{2}\tilde{E}_ {22}^{(3)}+\sqrt{3}\tilde{E}_ {8}^{(3)},\;
\tilde{E}_ {75}^{(3)}=2\sqrt{3}\tilde{E}_ {3}^{(3)}+2\sqrt{3}\tilde{E}_ {8}^{(3)}+\sqrt{3}\tilde{E}_ {22}^{(3)},\;
\tilde{E}_ {77}^{(3)}=2\sqrt{3}\tilde{E}_ {5}^{(3)}+2\sqrt{3}\tilde{E}_ {7}^{(3)},\;
\tilde{E}_ {78}^{(3)}=2\sqrt{3}\tilde{E}_ {6}^{(3)}+2\sqrt{3}\tilde{E}_ {8}^{(3)}+\sqrt{3}\tilde{E}_ {22}^{(3)},\;
\tilde{E}_ {82}^{(3)}=-2\sqrt{3}\tilde{E}_ {8}^{(3)}-3\sqrt{3}\tilde{E}_ {22}^{(3)},\;
\tilde{E}_ {85}^{(3)}=-\sqrt{3}\tilde{E}_ {11}^{(3)}+\sqrt{3}\tilde{E}_ {13}^{(3)},\;
\tilde{E}_ {86}^{(3)}=-\frac{3\sqrt{3}}{2}\tilde{E}_ {15}^{(3)}-\frac{\sqrt{3}}{2}\tilde{E}_ {19}^{(3)}+\sqrt{3}\tilde{E}_ {16}^{(3)},\;
\tilde{E}_ {89}^{(3)}=2\sqrt{3}\tilde{E}_ {11}^{(3)}-2\sqrt{3}\tilde{E}_ {13}^{(3)}-4\sqrt{3}\tilde{E}_ {12}^{(3)},\;
\tilde{E}_ {91}^{(3)}=2\sqrt{3}\tilde{E}_ {16}^{(3)}-\sqrt{3}\tilde{E}_ {15}^{(3)}+\sqrt{3}\tilde{E}_ {19}^{(3)},\;
\tilde{E}_ {92}^{(3)}=2\sqrt{3}\tilde{E}_ {15}^{(3)}+2\sqrt{3}\tilde{E}_ {19}^{(3)}-4\sqrt{3}\tilde{E}_ {16}^{(3)},\;
\tilde{E}_ {94}^{(3)}=2\sqrt{3}\tilde{E}_ {16}^{(3)}-\sqrt{3}\tilde{E}_ {15}^{(3)}+\sqrt{3}\tilde{E}_ {19}^{(3)},\;
\tilde{E}_ {95}^{(3)}=-2\sqrt{3}\tilde{E}_ {16}^{(3)}+3\sqrt{3}\tilde{E}_ {19}^{(3)}+\sqrt{3}\tilde{E}_ {15}^{(3)},\;
\tilde{E}_ {96}^{(3)}=2\sqrt{3}\tilde{E}_ {15}^{(3)}-2\sqrt{3}\tilde{E}_ {19}^{(3)}-4\sqrt{3}\tilde{E}_ {16}^{(3)}.
\end{autobreak}
\end{align}

\begin{align}\label{eqA6}
\begin{autobreak}
~
\tilde{E}_ {1}^{(3)}=\frac{1}{12}\tilde{C}_ {19}^{(3)}-\frac{1}{4}\tilde{C}_ {17}^{(3)},\;
\tilde{E}_ {2}^{(3)}=\frac{1}{12}\tilde{C}_ {1}^{(3)}+\frac{13}{24}\tilde{C}_ {19}^{(3)}+\frac{1}{6}\tilde{C}_ {10}^{(3)}+\frac{1}{6}\tilde{C}_ {13}^{(3)}-\frac{5}{8}\tilde{C}_ {17}^{(3)},\;
\tilde{E}_ {3}^{(3)}=\frac{1}{2}\tilde{C}_ {17}^{(3)}+\frac{1}{2}\tilde{C}_ {18}^{(3)}-\frac{1}{6}\tilde{C}_ {20}^{(3)},\;
\tilde{E}_ {4}^{(3)}=-\frac{1}{12}\tilde{C}_ {1}^{(3)}-\frac{1}{6}\tilde{C}_ {10}^{(3)}-\frac{1}{6}\tilde{C}_ {11}^{(3)}-\frac{1}{6}\tilde{C}_ {15}^{(3)}+\frac{1}{6}\tilde{C}_ {20}^{(3)}-\frac{3}{2}\tilde{C}_ {17}^{(3)}-\frac{3}{2}\tilde{C}_ {18}^{(3)},\;
\tilde{E}_ {5}^{(3)}=-\frac{1}{6}\tilde{C}_ {17}^{(3)}+\frac{1}{6}\tilde{C}_ {19}^{(3)},\;
\tilde{E}_ {6}^{(3)}=\frac{1}{3}\tilde{C}_ {17}^{(3)}+\frac{1}{3}\tilde{C}_ {18}^{(3)},\;
\tilde{E}_ {7}^{(3)}=\frac{1}{12}\tilde{C}_ {1}^{(3)}+\frac{1}{6}\tilde{C}_ {10}^{(3)}-\frac{1}{6}\tilde{C}_ {13}^{(3)}+\frac{1}{6}\tilde{C}_ {17}^{(3)}-\frac{1}{6}\tilde{C}_ {19}^{(3)},\;
\tilde{E}_ {8}^{(3)}=\frac{1}{12}\tilde{C}_ {1}^{(3)}-\frac{1}{3}\tilde{C}_ {17}^{(3)}-\frac{1}{3}\tilde{C}_ {18}^{(3)}+\frac{1}{6}\tilde{C}_ {10}^{(3)}+\frac{1}{6}\tilde{C}_ {11}^{(3)}-\frac{1}{6}\tilde{C}_ {15}^{(3)},\;
\tilde{E}_ {34}^{(3)}=\frac{11}{2}\tilde{C}_ {17}^{(3)}+\frac{11}{2}\tilde{C}_ {18}^{(3)}-\frac{1}{2}\tilde{C}_ {1}^{(3)}-\frac{1}{2}\tilde{C}_ {20}^{(3)}+\tilde{C}_ {15}^{(3)},\;
\tilde{E}_ {35}^{(3)}=-9\tilde{C}_ {17}^{(3)}-9\tilde{C}_ {18}^{(3)}-\tilde{C}_ {10}^{(3)}-\tilde{C}_ {11}^{(3)}-\tilde{C}_ {1}^{(3)}-\tilde{C}_ {15}^{(3)}+\tilde{C}_ {20}^{(3)},\;
\tilde{E}_ {36}^{(3)}=\frac{1}{2}\tilde{C}_ {10}^{(3)}+\frac{1}{2}\tilde{C}_ {11}^{(3)}+\frac{1}{2}\tilde{C}_ {15}^{(3)}-\frac{1}{2}\tilde{C}_ {20}^{(3)}+\frac{1}{4}\tilde{C}_ {1}^{(3)}+\frac{1}{4}\tilde{C}_ {19}^{(3)}+\frac{15}{4}\tilde{C}_ {17}^{(3)}+\frac{9}{2}\tilde{C}_ {18}^{(3)},\;
\tilde{E}_ {37}^{(3)}=-\frac{11}{2}\tilde{C}_ {18}^{(3)}+\frac{1}{2}\tilde{C}_ {10}^{(3)}+\frac{1}{2}\tilde{C}_ {13}^{(3)}+\frac{1}{2}\tilde{C}_ {20}^{(3)}+\frac{13}{8}\tilde{C}_ {19}^{(3)}+\frac{1}{4}\tilde{C}_ {1}^{(3)}-\frac{59}{8}\tilde{C}_ {17}^{(3)}-\tilde{C}_ {15}^{(3)},\;
\tilde{E}_ {38}^{(3)}=-4\tilde{C}_ {17}^{(3)}-4\tilde{C}_ {18}^{(3)}-\tilde{C}_ {15}^{(3)},\;
\tilde{E}_ {39}^{(3)}=-\frac{1}{3}\tilde{C}_ {10}^{(3)}-\frac{1}{3}\tilde{C}_ {11}^{(3)}+\frac{1}{3}\tilde{C}_ {15}^{(3)}-\frac{2}{3}\tilde{C}_ {1}^{(3)}+\frac{2}{3}\tilde{C}_ {17}^{(3)}+\frac{2}{3}\tilde{C}_ {18}^{(3)}+\frac{3}{2}\tilde{C}_ {2}^{(3)},\;
\tilde{E}_ {40}^{(3)}=-\frac{1}{3}\tilde{C}_ {10}^{(3)}-\frac{1}{3}\tilde{C}_ {11}^{(3)}+\frac{1}{3}\tilde{C}_ {15}^{(3)}-\frac{2}{3}\tilde{C}_ {1}^{(3)}+\frac{2}{3}\tilde{C}_ {17}^{(3)}+\frac{2}{3}\tilde{C}_ {18}^{(3)}-\frac{3}{2}\tilde{C}_ {2}^{(3)},\;
\tilde{E}_ {41}^{(3)}=\frac{1}{2}\tilde{C}_ {19}^{(3)}+\frac{1}{3}\tilde{C}_ {10}^{(3)}+\frac{1}{3}\tilde{C}_ {11}^{(3)}-\frac{1}{3}\tilde{C}_ {15}^{(3)}+\frac{1}{6}\tilde{C}_ {1}^{(3)}-\frac{2}{3}\tilde{C}_ {18}^{(3)}-\frac{7}{6}\tilde{C}_ {17}^{(3)},\;
\tilde{E}_ {42}^{(3)}=\frac{1}{3}\tilde{C}_ {10}^{(3)}+\frac{1}{3}\tilde{C}_ {11}^{(3)}-\frac{1}{3}\tilde{C}_ {15}^{(3)}+\frac{1}{3}\tilde{C}_ {17}^{(3)}+\frac{1}{3}\tilde{C}_ {18}^{(3)}+\frac{1}{6}\tilde{C}_ {1}^{(3)},\;
\tilde{E}_ {44}^{(3)}=\frac{1}{2}\tilde{C}_ {10}^{(3)}+\frac{1}{2}\tilde{C}_ {11}^{(3)}-\frac{1}{2}\tilde{C}_ {15}^{(3)}+\frac{1}{4}\tilde{C}_ {1}^{(3)}-\tilde{C}_ {17}^{(3)}-\tilde{C}_ {18}^{(3)},\;
\tilde{E}_ {45}^{(3)}=\frac{1}{2}\tilde{C}_ {10}^{(3)}+\frac{1}{2}\tilde{C}_ {11}^{(3)}-\frac{1}{2}\tilde{C}_ {15}^{(3)}+\frac{5}{4}\tilde{C}_ {1}^{(3)}-\tilde{C}_ {17}^{(3)}-\tilde{C}_ {18}^{(3)},\;
\tilde{E}_ {46}^{(3)}=-\frac{1}{2}\tilde{C}_ {11}^{(3)}-\frac{1}{2}\tilde{C}_ {13}^{(3)}+\frac{1}{2}\tilde{C}_ {15}^{(3)}-\frac{1}{2}\tilde{C}_ {19}^{(3)}+\frac{3}{2}\tilde{C}_ {17}^{(3)}+\tilde{C}_ {18}^{(3)},\;
\tilde{E}_ {47}^{(3)}=-\frac{1}{2}\tilde{C}_ {30}^{(3)}+\frac{1}{2}\tilde{C}_ {32}^{(3)},\;
\tilde{E}_ {48}^{(3)}=-\frac{1}{2}\tilde{C}_ {31}^{(3)}+\tilde{C}_ {32}^{(3)},\;
\tilde{E}_ {49}^{(3)}=-\frac{1}{2}\tilde{C}_ {21}^{(3)}+\frac{1}{2}\tilde{C}_ {25}^{(3)},\;
\tilde{E}_ {50}^{(3)}=-\frac{1}{2}\tilde{C}_ {22}^{(3)}+\tilde{C}_ {25}^{(3)},\;
\tilde{E}_ {51}^{(3)}=-\frac{1}{2}\tilde{C}_ {21}^{(3)}+\frac{1}{2}\tilde{C}_ {25}^{(3)},\;
\tilde{E}_ {60}^{(3)}=-\tilde{C}_ {37}^{(3)}+\tilde{C}_ {6}^{(3)},\;
\tilde{E}_ {62}^{(3)}=\tilde{C}_ {37}^{(3)}+\tilde{C}_ {6}^{(3)},\;
\tilde{E}_ {63}^{(3)}=\frac{1}{2}\tilde{C}_ {39}^{(3)}+\tilde{C}_ {37}^{(3)},\;
\tilde{E}_ {72}^{(3)}=\frac{1}{2\sqrt{3}}\tilde{C}_ {10}^{(3)}-\frac{1}{2\sqrt{3}}\tilde{C}_ {13}^{(3)}+\frac{1}{4\sqrt{3}}\tilde{C}_ {1}^{(3)}-\frac{1}{\sqrt{3}}\tilde{C}_ {17}^{(3)},\;
\tilde{E}_ {73}^{(3)}=-\frac{11}{4\sqrt{3}}\tilde{C}_ {17}^{(3)}+\frac{1}{\sqrt{3}}\tilde{C}_ {1}^{(3)}+\frac{2}{\sqrt{3}}\tilde{C}_ {10}^{(3)}+\frac{3\sqrt{3}}{4}\tilde{C}_ {19}^{(3)},\;
\tilde{E}_ {74}^{(3)}=-\frac{10}{\sqrt{3}}\tilde{C}_ {17}^{(3)}-\frac{10}{\sqrt{3}}\tilde{C}_ {18}^{(3)}-\frac{1}{2\sqrt{3}}\tilde{C}_ {10}^{(3)}-\frac{1}{2\sqrt{3}}\tilde{C}_ {11}^{(3)}+\frac{1}{2\sqrt{3}}\tilde{C}_ {37}^{(3)}-\frac{1}{4\sqrt{3}}\tilde{C}_ {1}^{(3)}+\frac{1}{\sqrt{3}}\tilde{C}_ {20}^{(3)}-\frac{\sqrt{3}}{2}\tilde{C}_ {15}^{(3)},\;
\tilde{E}_ {75}^{(3)}=\frac{1}{2\sqrt{3}}\tilde{C}_ {1}^{(3)}+\frac{1}{\sqrt{3}}\tilde{C}_ {10}^{(3)}+\frac{1}{\sqrt{3}}\tilde{C}_ {11}^{(3)}-\frac{1}{\sqrt{3}}\tilde{C}_ {15}^{(3)}+\frac{1}{\sqrt{3}}\tilde{C}_ {17}^{(3)}+\frac{1}{\sqrt{3}}\tilde{C}_ {18}^{(3)}-\frac{1}{\sqrt{3}}\tilde{C}_ {20}^{(3)}+\frac{1}{\sqrt{3}}\tilde{C}_ {37}^{(3)},\;
\tilde{E}_ {77}^{(3)}=\frac{1}{2\sqrt{3}}\tilde{C}_ {1}^{(3)}+\frac{1}{\sqrt{3}}\tilde{C}_ {10}^{(3)}-\frac{1}{\sqrt{3}}\tilde{C}_ {13}^{(3)},\;
\tilde{E}_ {78}^{(3)}=\frac{1}{2\sqrt{3}}\tilde{C}_ {1}^{(3)}+\frac{1}{\sqrt{3}}\tilde{C}_ {10}^{(3)}+\frac{1}{\sqrt{3}}\tilde{C}_ {11}^{(3)}-\frac{1}{\sqrt{3}}\tilde{C}_ {15}^{(3)}+\frac{1}{\sqrt{3}}\tilde{C}_ {37}^{(3)},\;
\tilde{E}_ {80}^{(3)}=-\frac{1}{2\sqrt{3}}\tilde{C}_ {1}^{(3)}-\frac{1}{\sqrt{3}}\tilde{C}_ {10}^{(3)}+\frac{1}{\sqrt{3}}\tilde{C}_ {13}^{(3)}-\frac{1}{\sqrt{3}}\tilde{C}_ {17}^{(3)}+\frac{1}{\sqrt{3}}\tilde{C}_ {19}^{(3)},\;
\tilde{E}_ {82}^{(3)}=-\frac{1}{2\sqrt{3}}\tilde{C}_ {1}^{(3)}-\frac{1}{\sqrt{3}}\tilde{C}_ {10}^{(3)}-\frac{1}{\sqrt{3}}\tilde{C}_ {11}^{(3)}+\frac{1}{\sqrt{3}}\tilde{C}_ {15}^{(3)}+\frac{2}{\sqrt{3}}\tilde{C}_ {17}^{(3)}+\frac{2}{\sqrt{3}}\tilde{C}_ {18}^{(3)}-\sqrt{3}\tilde{C}_ {37}^{(3)},\;
\tilde{E}_ {85}^{(3)}=\frac{1}{2\sqrt{3}}\tilde{C}_ {30}^{(3)}+\frac{1}{2\sqrt{3}}\tilde{C}_ {32}^{(3)},\;
\tilde{E}_ {86}^{(3)}=\frac{1}{4\sqrt{3}}\tilde{C}_ {22}^{(3)}-\frac{1}{4\sqrt{3}}\tilde{C}_ {25}^{(3)}+\frac{\sqrt{3}}{4}\tilde{C}_ {21}^{(3)},\;
\tilde{E}_ {89}^{(3)}=-\frac{1}{\sqrt{3}}\tilde{C}_ {30}^{(3)}-\frac{1}{\sqrt{3}}\tilde{C}_ {31}^{(3)}-\frac{1}{\sqrt{3}}\tilde{C}_ {32}^{(3)},\;
\tilde{E}_ {91}^{(3)}=\frac{1}{2\sqrt{3}}\tilde{C}_ {21}^{(3)}+\frac{1}{2\sqrt{3}}\tilde{C}_ {22}^{(3)}+\frac{1}{2\sqrt{3}}\tilde{C}_ {25}^{(3)},\;
\tilde{E}_ {92}^{(3)}=-\frac{1}{\sqrt{3}}\tilde{C}_ {21}^{(3)}-\frac{1}{\sqrt{3}}\tilde{C}_ {22}^{(3)}+\frac{1}{\sqrt{3}}\tilde{C}_ {25}^{(3)},\;
\tilde{E}_ {94}^{(3)}=\frac{1}{2\sqrt{3}}\tilde{C}_ {21}^{(3)}+\frac{1}{2\sqrt{3}}\tilde{C}_ {22}^{(3)}+\frac{1}{2\sqrt{3}}\tilde{C}_ {25}^{(3)},\;
\tilde{E}_ {95}^{(3)}=-\frac{1}{2\sqrt{3}}\tilde{C}_ {21}^{(3)}-\frac{1}{2\sqrt{3}}\tilde{C}_ {22}^{(3)}+\frac{\sqrt{3}}{2}\tilde{C}_ {25}^{(3)},\;
\tilde{E}_ {96}^{(3)}=-\frac{1}{\sqrt{3}}\tilde{C}_ {21}^{(3)}-\frac{1}{\sqrt{3}}\tilde{C}_ {22}^{(3)}-\frac{1}{\sqrt{3}}\tilde{C}_ {25}^{(3)}.
\end{autobreak}
\end{align}

\section{$\mathcal{O}(p^{4})$ order results}\label{p4r}
This appendix gives Tables \ref{tab:table9} and \ref{tab:table8}.
\begin{longtable}{lcrcrlcrcr}
\caption{Independent terms in the relativistic Lagrangian at the order $\mathcal{O}(p^{4})$. The columns 2 and 7 (4 and 9) are for the two (three)-flavor case. The columns 2, 4, 7, and 9 label the number for each term. The terms without a number are not independent. ``I'' means that the term is chosen as an independent term in the heavy diquark limit.
}\label{tab:table9}
\\
%\begin{ruledtabular}
\hline $O_n/o_n$  & $SU(2)$ &$\tilde{e}_{n}^{(4)}$&$SU(3)$&$\tilde{E}_{n}^{(4)}$&$O_n$  & $SU(2)$ &$\tilde{e}_{n}^{(4)}$&$SU(3)$&$\tilde{E}_{n}^{(4)}$\\
\hline%\addlinespace[2pt]
\endfirsthead
\hline$O_n/o_n$  & $SU(2)$ &$\tilde{e}_{n}^{(4)}$&$SU(3)$&$\tilde{E}_{n}^{(4)}$&$O_n$  & $SU(2)$ &$\tilde{e}_{n}^{(4)}$&$SU(3)$&$\tilde{E}_{n}^{(4)}$\\
\hline%\addlinespace[2pt]
\endhead
\hline
\endfoot
\hline
\endlastfoot

$\Bb\la u^{\mu}u_{\mu}\ra\la u^{\nu}u_{\nu}\ra B$ & 1& $\mathrm{I}\quad$ & 1& $\mathrm{I}\quad$ & $\Tb^{\mu}\la{f_{+}}^{\nu\lambda}\ra\la f_{+\nu\lambda}\ra T_{\mu}$ &  &  & 436&  \\
$\Bb\la u^{\mu}u^{\nu}\ra\la u_{\mu}u_{\nu}\ra B$ & 2& $\mathrm{I}\quad$ & 2& $\mathrm{I}\quad$ & $\Tb^{\mu}\la{f_{+\mu}}^{\nu}\ra\la{f_{+}}^{\lambda\rho}\ra D_{\nu\lambda}T_{\rho}$ &  &  & 437&  \\
$\Bb\la u^{\mu}u_{\mu}\ra\la u^{\nu}u^{\lambda}\ra D_{\nu\lambda}B$ & 3& $\mathrm{I}\quad$ & 3& $\mathrm{I}\quad$ & $\Tb^{\mu}\la{f_{+}}^{\nu\lambda}\ra\la{f_{+\nu}}^{\rho}\ra D_{\lambda\rho}T_{\mu}$ &  &  & 438&  \\
$\Bb\la u^{\mu}u^{\nu}\ra\la u_{\mu}u^{\lambda}\ra D_{\nu\lambda}B$ & 4& $\mathrm{I}\quad$ & 4& $\mathrm{I}\quad$ & $i\Tb^{\mu}\la{f_{+\mu}}^{\nu}\ra\la{f_{+}}^{\lambda\rho}\ra\sigma_{\nu\lambda}T_{\rho}$ &  &  & 439&  \\
$\Bb\la u^{\mu}u^{\nu}\ra\la u^{\lambda}u^{\rho}\ra D_{\mu\nu\lambda\rho}B$ & 5& $\mathrm{I}\quad$ & 5& $\mathrm{I}\quad$ & $\Tb^{\mu}\la{f_{+\mu}}^{\nu}\ra{f_{+\nu}}^{\lambda}T_{\lambda}+\mathrm{H.c.}$ & 213&  & 440&  \\
$\Bb\la u^{\mu}u_{\mu}\ra u^{\nu}u_{\nu}B$ &  &  & 6& $\mathrm{I}\quad$ & $\Tb^{\mu}\la{f_{+}}^{\nu\lambda}\ra f_{+\nu\lambda}T_{\mu}$ & 214&  & 441&  \\
$\Bb\la u^{\mu}u^{\nu}\ra u_{\mu}u_{\nu}B$ &  &  & 7& $\mathrm{I}\quad$ & $\Tb^{\mu}\la{f_{+\mu}}^{\nu}\ra{f_{+}}^{\lambda\rho}D_{\nu\lambda}T_{\rho}+\mathrm{H.c.}$ & 215&  & 442&  \\
$\Bb\la u^{\mu}u_{\mu}\ra u^{\nu}u^{\lambda}D_{\nu\lambda}B$ &  &  & 8& $\mathrm{I}\quad$ & $\Tb^{\mu}\la{f_{+}}^{\nu\lambda}\ra{f_{+\nu}}^{\rho}D_{\lambda\rho}T_{\mu}$ & 216&  & 443&  \\
\end{longtable}

\begin{longtable}{lcclcc}
\caption{Independent Lagrangian terms in the heavy diquark limit at the order $\mathcal{O}(p^{4})$. The second and fifth (third and sixth) columns label the number for each term in the two (three)-flavor case. The terms without a number are not independent.}\label{tab:table8}
\\
\hline\hline $P_n/p_n$  & $SU(2)$ & $SU(3)$&$P_n/p_n$  & $SU(2)$ & $SU(3)$\\
\hline%\addlinespace[2pt]
\endfirsthead
\hline\hline $P_n/p_n$  & $SU(2)$ & $SU(3)$&$P_n/p_n$  & $SU(2)$ & $SU(3)$\\
\hline%\addlinespace[2pt]
\endhead
\hline\hline
\endfoot
\hline\hline
\endlastfoot

$\psib^{\mu}\la u^{\nu}u_{\nu}\ra\la u^{\lambda}u_{\lambda}\ra\psi_{\mu}$ & 1 & 1 & $\epsilon^{\mu\nu\lambda\rho}\psib^{\sigma}\la u_{\mu}f_{-\nu\lambda}\ra u_{\rho}\psi_{\sigma}$ & 46 & 99  \\
$\psib^{\mu}\la u^{\nu}u^{\lambda}\ra\la u_{\nu}u_{\lambda}\ra\psi_{\mu}$ & 2 & 2 & $\epsilon^{\mu\nu\lambda\rho}\psib^{\sigma}\la u_{\mu}f_{-\nu\lambda}\ra u^{\delta}v_{\rho}v_{\delta}\psi_{\sigma}$ & 47 & 100  \\
$\psib^{\mu}\la u^{\nu}u_{\nu}\ra\la u^{\lambda}u^{\rho}\ra v_{\lambda}v_{\rho}\psi_{\mu}$ & 3 & 3 & $\epsilon^{\mu\nu\lambda\rho}\psib^{\sigma}\la u_{\mu}{f_{-\nu}}^{\delta}\ra u_{\lambda}v_{\rho}v_{\delta}\psi_{\sigma}$ & 48 & 101  \\
$\psib^{\mu}\la u^{\nu}u^{\lambda}\ra\la u_{\nu}u^{\rho}\ra v_{\lambda}v_{\rho}\psi_{\mu}$ & 4 & 4 & $\psib^{\mu}\la{f_{+}}^{\nu\lambda}\ra{f_{-\nu}}^{\rho}\gamf\gamma_{\lambda}v_{\rho}\psi_{\mu}$ & 49 &  \\
$\psib^{\mu}\la u^{\nu}u^{\lambda}\ra\la u^{\rho}u^{\sigma}\ra v_{\nu}v_{\lambda}v_{\rho}v_{\sigma}\psi_{\mu}$ & 5 & 5 & $\psib^{\mu}\la{f_{+}}^{\nu\lambda}\ra{f_{-\nu}}^{\rho}\gamf\gamma_{\rho}v_{\lambda}\psi_{\mu}$ & 50 &  \\
$\psib^{\mu}\la u^{\nu}u_{\nu}\ra u^{\lambda}u_{\lambda}\psi_{\mu}$ &  & 6 & $\epsilon^{\mu\nu\lambda\rho}\psib^{\sigma}u_{\mu}u_{\nu}f_{-\lambda\rho}\psi_{\sigma}+\mathrm{H.c.}$ &  & 102  \\
$\psib^{\mu}\la u^{\nu}u^{\lambda}\ra u_{\nu}u_{\lambda}\psi_{\mu}$ &  & 7 & $\epsilon^{\mu\nu\lambda\rho}\psib^{\sigma}u_{\mu}u_{\nu}{f_{-\lambda}}^{\delta}v_{\rho}v_{\delta}\psi_{\sigma}+\mathrm{H.c.}$ &  & 103  \\
$\psib^{\mu}\la u^{\nu}u_{\nu}\ra u^{\lambda}u^{\rho}v_{\lambda}v_{\rho}\psi_{\mu}$ &  & 8 & $\epsilon^{\mu\nu\lambda\rho}\psib^{\sigma}u_{\mu}u^{\delta}f_{-\nu\lambda}v_{\rho}v_{\delta}\psi_{\sigma}+\mathrm{H.c.}$ &  & 104  \\
\end{longtable}

%\input{latex(US).tex}
%\end{document}

\bibliography{bibtex}% Produces the bibliography via BibTeX.

\end{document}